\documentstyle[12pt,epsf]{article}
\newcommand{\beq}{\begin{equation}}
\newcommand{\eeq}{\end{equation}}
\newcommand{\beqa}{\begin{eqnarray}}
\newcommand{\eeqa}{\end{eqnarray}}

\newcommand{\ve}{\varepsilon}

\newcommand{\qp}{\tilde{Q}_+}
\newcommand{\qm}{\tilde{Q}_-}
\newcommand{\na}{\nabla}
\newcommand{\no}{\nonumber}
\newcommand{\pp}{M_{\pi^+}}
\newcommand{\pps}{M_{\pi^+}^2}
\newcommand{\pn}{M_{\pi^0}}
\newcommand{\pns}{M_{\pi^0}^2}
\newcommand{\sig}{\sigma^{\mu\nu}}
\newcommand{\Sig}{v_\rho \ve^{\rho\mu\nu\sigma} \, S_\sigma}
\newcommand{\SIG}{ i v^\mu S^\nu }
\newcommand{\hc}{ + h.c. }

\newcommand{\vs}{\vspace{-0.25cm}}
\newcommand{\fps}{F_\pi^2}
\newcommand{\llo}{\ln\frac{M_{\pi^0}}{\lambda}}
\newcommand{\lm}{\ln\frac{M_{\pi^+}}{M_{\pi^0}}}

\begin{document}

\begin{flushright}
{\small
FZJ-IKP(TH)-1999-03 \\ 
UWThPh-1999-15 }
\end{flushright}


\vspace{2cm}

\begin{center}

{{\Large\bf 
Virtual photons in baryon chiral perturbation theory\footnote{Work 
supported in part by TMR, EC-Contract No. ERBFMRX-CT980169 (EURODA$\Phi$NE).}  
}}

\end{center}

\vspace{.4in}

\begin{center}
{\large
Guido M\"uller$^\star$\footnote{email:
gmueller@doppler.thp.univie.ac.at}, $\,$
Ulf-G. Mei{\ss}ner$^\ddagger$\footnote{email: Ulf-G.Meissner@fz-juelich.de}}

\bigskip

$^\star${\it Universit\"at Wien, Institut f\"ur Theoretische Physik\\
Boltzmanngasse 5, A--1090 Wien, Austria}

\bigskip

$^\ddagger${\it Forschungszentrum J\"ulich, Institut f\"ur Kernphysik
(Theorie)\\ D-52425 J\"ulich, Germany}

\bigskip

\end{center}

\vspace{.7in}

\thispagestyle{empty}

\begin{abstract}
\noindent
We construct the general Lagrangian for relativistic and heavy baryon
chiral perturbation theory with virtual photons to fourth order. We
work out the electromagnetic and strong isospin violating
contributions to the nucleon self--energy, the nucleon mass and
the scalar form factor of the nucleon. Electromagnetic effects for
the shift to the Cheng--Dashen point can be as large as 2~MeV.
We also discuss the corrections
to Weinberg's prediction for the scattering length difference $a(\pi^0
p) - a(\pi^0 n)$ and show that they are small.
\end{abstract}

\vspace{1in}


\vfill

\pagebreak

\section{Introduction}

There has been renewed interest in precisely calculating electromagnetic (em)
corrections  to low energy strong and semi--leptonic processes involving
the (pseudo) Goldstone bosons of the strong interactions. For example,
the two loop calculations for elastic pion--pion scattering in the framework
of chiral perturbation theory, which is the effective field theory (EFT) of
the Standard Model at low energies, have now reached such a precision that 
it is mandatory to also evaluate the pertinent dual effects of virtual photons.
Furthermore, to systematically investigate the violation of isospin symmetry 
one has to account for its two competing sources consistently, namely the 
light quark mass difference $m_u - m_d$ as well as the virtual photon effects. 
Efforts to include virtual photons in the mesonic effective chiral Lagrangian 
to and beyond leading order based on an extended power counting and
to calculate consequences for a variety of processes can be found e.g. in 
refs.\cite{gepdR}-\cite{KU}. 

Baryon chiral perturbation theory offers another possibility of investigating
isospin violation. As first stressed by Weinberg~\cite{weinmass}, reactions
involving nucleons and neutral pions can lead to gross violations of isospin,
e.g. in the scattering length difference $a(\pi^0 p) - a(\pi^0 n)$
he predicted an effect of the order of 30\%. This is
because chiral symmetry and isospin breaking appear  at the same order
and the leading isospin symmetric terms involving neutral pions are suppressed
due to chiral symmetry. This calculation was sharpened and extended also
to the so--called pion--nucleon $\sigma$--term in ref.\cite{ms} (which we
call ``MS'' from here on). In that paper the complete effective chiral 
pion--nucleon Lagrangian  in the presence of virtual photons to one loop and third
order in small momenta was  constructed. These considerations were extended
to other threshold pion--nucleon channels in~\cite{FMSi}. It is, however, known 
that precise and complete one loop calculations in the baryon sector should
be carried out to fourth order for two reasons. First, the chiral
expansion in small momenta
and meson masses proceeds in steps of one power in the presence of baryons
(unlike in the meson sector, were only even powers in small momenta or
meson masses are
allowed for a large quark condensate), so that a complete one loop calculation
should include terms of orders $q^3$ and $q^4$ (where $q$ collectively
denotes the small expansion parameters). Second, it has also been shown that
in many cases one loop graphs with exactly one dimension two insertion are
fairly large (which is related to the fact that the corresponding coupling
constants encode the leading effects due
to the close-by $\Delta(1232)$ resonance~\cite{bkmlec}). 
These terms are of ${\cal O}(q^4)$.
It has also been argued that the chiral expansions in odd and even powers
should be considered separately, i.e. a one loop calculation to third order
only gives the leading corrections to the dimension one tree graphs.

Most calculations in baryon CHPT are performed in the so--called heavy
fermion formulation~\cite{jm,bkkm}, which is called 
heavy baryon chiral perturbation theory (HBCHPT). 
This is based on the observation
that a straightforward extension of CHPT with baryons treated fully relativistically
leads to a considerable complication in the power counting since only
nucleon three--momenta are small compared to typical hadronic scales,
as discussed in detail in ref.\cite{gss}.  Another advantage of the
heavy fermion formulation is its extreme computational simplicity, since
all Dirac $\gamma$--matrices are expressed in terms of the nucleons 
four--velocity and
spin vector. However, one has to be careful with strict non--relativistic
expansions since in some cases they can conflict strictures from analyticity,
as discussed e.g. in ref.\cite{bkmff}. Therefore, in ref.\cite{BL}, a novel
scheme was proposed which is based on the relativistic formulation but
circumvents the power counting problems (to one loop) 
by a clever separation of the loop integrals into IR singular and regular parts.
In this formulation, all analytic constraints are fulfilled by construction.
We remark, however, that the calculational simplicity of the heavy fermion
approach is lost. For comparison, neutral pion photoproduction off nucleons
calculated relativistically
involves some 63 loop graphs~\cite{bkmrpi0} to ${\cal O}(q^3)$ but only
4 to the same order in HBCHPT~\cite{bkmhpi0}. For these reasons, we give
the effective Lagrangian with virtual photons in the relativistic as well as
in the heavy baryon formulation. In any case, in a systematic construction
of the generating functional for HBCHPT, one should start from the fully
relativistic theory~\cite{bkkm}.

The manuscript is organized as follows. In section~2,
we construct the fourth order pion--nucleon Lagrangian with
virtual photons. First, we write down the pertinent building blocks
and construction principles. Then, we summarize what is known about
the first three orders of the effective Lagrangian and then give
explicitely the fourth order terms, both in the relativistic and the
heavy baryon formulation. Some applications are discussed in
section~3. First, we consider the nucleon self--energy and mass shift
including also photon loops (as demanded by consistency).
To deal with the IR divergences stemming from the masslessness of the
photon, we introduce a small photon mass $m_\gamma$ and additional 
``photonic'' counterterms $\sim m_\gamma^2$. Their effect cancels in
physical observables but these terms need to be retained in
intermediate steps. We then consider the isospin--violating
corrections to the nucleon scalar form factor $\sigma_{\pi N} (t)$, 
more precisely we are concerned with the shift to the Cheng--Dashen point,  
$\sigma_{\pi N} (2M_{\pi^0}^2) - \sigma_{\pi N} (0)$. As a third
application, we work out the corrections to Weinberg's famous 
prediction concerning the S--wave scattering lengths difference
for neutral pions off
nucleons, $a(\pi^0 p) - a(\pi^0 n)$, which to third
order is given entirely in terms of the light quark mass
difference $m_u -m_d$. 
Section~4 contains the summary and conclusions. Some more technical
material is relegated to the appendices.
 
\section{Construction of the effective Lagrangian}
\label{sec:L}
\def\theequation{\arabic{section}.\arabic{equation}}
\setcounter{equation}{0}

In this section, we will be concerned with the construction of the 
complete 
 fourth order effective Lagrangian for
two flavor baryon chiral perturbation theory. We give its relativistic
and heavy baryon formulation for the reasons discussed in the
introduction. Since the details of the construction are similar to the
ones exposed  in ref.\cite{FMS},\footnote{The third order
effective pion-nucleon Lagrangian was first given in another basis
in~\cite{eckmoj}.}
we concentrate on discussing the new aspects due to the inclusion of 
the virtual photons. While our considerations are general, we give
specific details only for the two flavor case of spontaneously broken
chiral SU(2)$_L \times$SU(2)$_R$ symmetry.

\subsection{Single--nucleon effective field theory and building blocks}
\label{sec:HBCHPT}

The starting point of our approach is to construct the most general
chiral invariant Lagrangian built from pions, nucleons and external
scalar, pseudoscalar, vector, axial--vector sources {\it and}
virtual photons, parametrized in terms of the vector field $A_\mu
(x)$. The Goldstone
bosons are collected in a 2$\times$2 matrix-valued field $U(x)= u^2(x)$.
The nucleons are described by structureless relativistic spin-${\small
  \frac{1}{2}}$ particles, the spinors being denoted by $\Psi
(x)$ in the relativistic case 
or by the so--called light component $N(x)$ in the heavy fermion
formulation. The effective field theory admits a low energy expansion, i.e. the
corresponding effective Lagrangian can be written as 
\beq
{\cal L}_{\rm eff} = 
 {\cal L}_{\pi\pi}^{(2)} +  {\cal L}_{\pi\pi}^{(4)} +
 {\cal L}_{\pi N}^{(1)} +  {\cal L}_{\pi N}^{(2)} +
 {\cal L}_{\pi N}^{(3)} +  {\cal L}_{\pi N}^{(4)} + \ldots~,
\eeq
where the ellipsis denotes terms of higher order not considered
here. For the explicit
form of the meson Lagrangian and the dimension one and two
pion--nucleon terms, we refer to ref.\cite{bkmrev}. More precisely,
in the pion--nucleon sector,
the inclusion of the virtual photons modifies the leading
term of dimension one (as given below~\cite{ms}) and leads to new 
local (contact) terms starting at second order,
\beq
{\cal L}_{\pi N}^{(i)} = {\cal L}_{\pi N, \rm str}^{(i)}  
+ {\cal L}_{\pi N, \rm em}^{(i)}\, , \quad (i = 2,3,4)~.
\eeq
We are concerned here with the construction of the terms
including virtual photons ${\cal L}_{\pi N, \rm em}^{(i)}$ $(i=2,3,4)$.
The minimal form of the strong terms (no virtual photons) 
up-to-and-including third order  can be found in ref.\cite{FMS} and the
divergence structure of ${\cal L}_{\pi N, \rm str}^{(4)}$ is discussed
in ref.\cite{MMS}. Two groups are presently working on setting up the
minimal fourth order strong Lagrangian~\cite{FMMS,moj}.
We remark that the various parameters  appearing in the effective
Lagrangian have to be taken in the chiral SU(2) limit ($m_u=m_d=0\, ,
\,\, m_s$ fixed, $e^2 = 0$). The corresponding dimension two and four
electromagnetic meson Lagrangians are briefly reviewed in app.~\ref{app:LM}.

To construct the pertinent terms, we use the standard methods of non--linearly
realized chiral symmetry. From the external sources, we construct the
following building blocks starting with the nucleon covariant derivative 
$\nabla_\mu$, the chiral connection $\Gamma_\mu$, the chiral viel--bein $u_\mu$ and
the external sources: 
\beqa
\nabla_\mu & = & \partial_\mu\, + \Gamma_\mu\,~, \\ 
\Gamma_\mu & = & \frac{1}{2}\,[u^\dagger,\partial_\mu u] 
                 - \frac{i}{2}\,u^\dagger r_\mu u
                 - \frac{i}{2}\,u l_\mu u^\dagger ~,\\
u_\mu & = & i\,\left(\partial_\mu u\,u^\dagger + u^\dagger\partial_\mu u 
            -i\,u^\dagger r_\mu u + i\,u l_\mu u^\dagger\right)~, \\
r_\mu & = & v_\mu + a_\mu~, \quad l_\mu  =  v_\mu - a_\mu~, \\
F^\pm_{\mu\nu} & = & u^\dagger F^R_{\mu\nu}u \pm u F^L_{\mu\nu}u^\dagger~
,\\
F^L_{\mu\nu} & = & \partial_\mu l_\nu - \partial_\nu l_\mu - i\,[l_\mu,l_
\nu]~, \\
F^R_{\mu\nu} & = & \partial_\mu r_\nu - \partial_\nu r_\mu - i\,[r_\mu,r_
\nu]~, \\
\chi_\pm & = & u^\dagger \chi u^\dagger \pm u \chi^\dagger u~,
\eeqa
and we employ the following definition for traceless operators
('$\langle \ldots \rangle$' denotes the trace in flavor space)
\beq
\tilde{A}  =  A - \frac{1}{2}\,\langle A \rangle~. 
\eeq
Here and in what follows, $A$ denotes an arbitrary 2$\times$2 matrix.
Furthermore, $\chi (x)= s(x) + ip(x)$ includes the explicit chiral symmetry breaking
through the small current quark masses, $s(x)= B \, {\rm diag}(m_u,m_d) +
\ldots$ and $B = |\langle 0 | \bar q q | 0 \rangle | / F_\pi^2$,
with $F_\pi$ the (weak) pion decay constant and the scalar quark
condensate measures the strength of the chiral symmetry breaking.
We assign the following chiral dimensions to the pertinent
fields and operators:
\beqa
&& U(x) , \Psi (x) ,   \partial_\mu \Psi (x), N(x) = {\cal O}(1)~, \nonumber \\
&& \partial _\mu U(x) , \partial_\mu N(x) , u_\mu (x) , 
l_\mu (x) , r_\mu (x) = {\cal O}(q)~,\\
&& s(x), p(x), \chi_\pm (x), F_{\mu\nu}^\pm (x) ,   F_{\mu\nu}^{L,R} (x)  
= {\cal O}(q^2)~,\nonumber
\eeqa
which amounts to the so--called large condensate scenario of the
spontaneous chiral symmetry violation, $B \gg F_\pi$.
Here, $q$ denotes a genuine small momentum or meson mass with respect
to the typical hadronic scale of about  1~GeV.

To introduce virtual photons in the effective pion--nucleon field
theory, one has to first set up to a power counting scheme for
the electric charge $e$. In principle, one has to deal with two 
expansions, one in the small momenta/masses (the chiral expansion)
and a second one in the em coupling. A priori, these two expansions
can be treated separately.
As proposed in ref.\cite{urech} and based
on the observation that $e^2/4\pi \simeq M_\pi^2 / (4\pi F_\pi )^2
\sim 1/100$, we count the electric charge as a small momentum,
\beq 
e = {\cal O}(q)~,
\eeq
which is the most economic way of organizing the double expansions
in terms of one. In
particular, since the electric charge related to the virtual photons
always appears quadratic,
the following pattern for the terms in the electromagnetic effective
Lagrangian emerges.\footnote{Of course, the external vector field can
  be a photon. These are counted in the standard manner and do not 
  concern our arguments.} 
At second order, we can only have terms of order
$e^2$, at third order $e^2 q$ and at fourth order $e^2 q^2$ or $e^4$
(besides the standard strong terms).
In what follows, we will collectively denote a term of order $k$
as ${\cal O}(q^k)$ without differentiating between small momenta or em couplings.
Note that the necessity of having the charge matrix appearing squared
severely limits the number of possible terms. For example,
em corrections to the explicit symmetry breaking operators, which are of second
order in the standard counting, can therefore only appear
at fourth order. The inclusion of the virtual photons proceeds as follows.
In the meson sector one usually works with the quark charge matrix 
\beqa \label{qmes}
Q = {e \over 3} \left( \matrix { 2 & 0 \nonumber \\ 0 & -1
 \nonumber\\} \!\!\!\!\!\!\!\!\!\!\!\!\!\!\!\!\!\!  \right) 
= {e \over 6} \left( 3 \tau^3 + {\bf 1}  \right)
 \,\, 
\eeqa
and introduces spurions $Q_{L,R}$ with a definite transformation
property under chiral SU(2)$_L \times$ SU(2)$_R$,
\beq\label{Spurm}
Q_{I} \to g_I \, Q_{I} \, g_I^\dagger~, \quad g_I \in {\rm
  SU}(2)_I~, \quad I = L,R
\eeq 
to construct chiral invariant building blocks. Only at the end, one
sets $Q_L = Q_R = Q$~\cite{gepdR,urech}. 
In the coupled pion--nucleon sector it is natural to use the nucleon
charge matrix~\cite{NR,ms} 
\beqa \label{qnuc}
Q = {e} \left( \matrix { 1 & 0 \nonumber \\ 0 & 0
 \nonumber\\} \!\!\!\!\!\!\!\!\!\!\!\!\!\!\!\!\!\!  \right) 
= {e \over 2} \left(  \tau^3 + {\bf 1}  \right)
 \,\, ,
\eeqa
and to work with the usual generalizations,
\beq
Q_{\pm} = \frac{1}{2} \, \left(u \, Q \,u^\dagger \pm u^\dagger \, Q \, 
u\right) \,\, ,
\eeq
which under chiral SU(2)$_L\times$SU(2)$_R$ symmetry transform as 
any matrix--valued matter field
\beq
Q_\pm  \to K \, Q_\pm  \, K^\dagger~.
\eeq 
Here, $K$ is the compensator field representing an element of the conserved subgroup
SU(2)$_V$. It depends on the pion fields in a highly non--linear
fashion. Furthermore, under parity ($P$) and charge conjugation ($C$)
transformations, one finds
\beqa
P \, Q_\pm \, P^{-1} &=& \pm \, Q_\pm \,\,\, , \\ 
C \, Q_\pm \, C^{-1} &=& \pm \, Q_\pm^T \,\,\, , 
\eeqa
where $Q^T$ is the transposed of the matrix $Q$. 
The physical results do, of course, not depend on the specific
parametrization. From the decomposition 
\beqa
Q = e \left( \alpha {\bf 1}   + \beta \tau^3 \right)  
\eeqa
with $\alpha =1/6 , \beta = 1/2 $ in eq.(\ref{qmes}) and 
 $\alpha =1/2 , \beta = 1/2 $ in eq.(\ref{qnuc}) one
derives the relation
\beqa
\langle Q \rangle^{2}  &=& \frac{(2\alpha)^{2}}{\alpha^2 + \beta^2} 
\, \langle Q^2 \rangle
\eeqa 
which allows to drop the monomials proportional to 
$\langle Q \rangle^2$ and $\langle Q \rangle^4$ in 
the Lagrangian to fourth order. In what follows we give the minimal 
Lagrangian to the fourth order with virtual photons. For completeness
the Lagrangian of second and third order is also given. We use the 
notation of SM, but prefer to work in the flavor basis with traceless
objects, as in ref.\cite{FMS}. Finally, we remark that throughout
we work in the Landau gauge, see also app.~\ref{app:LM}.

\subsection{First, second and third order terms}
\label{sec:L123}
In this paragraph, we briefly review what is already known about the
electromagnetic effective pion--nucleon Lagrangian to third order,
following closely SM (with a few differences to be discussed).

In particular, to lowest order ($i=1$) one finds (in the relativistic and
the heavy fermion formulation)
\beqa
{\cal L}_{\pi N}^{(1)} &=& \bar{\Psi} \,\biggl( i \gamma_\mu \cdot 
\tilde{\nabla}^\mu -
m + \frac{1}{2}g_A \, \gamma^\mu \gamma_5 \cdot \tilde{u}_\mu \, \biggr) \, \Psi
\nonumber \\
&=& \bar{N} \,\biggl( i v \cdot \tilde{\nabla} +
g_A \, S \cdot \tilde{u} \, \biggr) \, N + {\cal O}(\frac{1}{m})\,\,\, , 
\eeqa
with 
\beqa \label{covder}
\tilde{\nabla}_\mu &=& \nabla_\mu -i \, Q_+ \, A_\mu \,\,\, ,  \\
\tilde{u}_\mu &=& u_\mu - 2 \, Q_- \, A_\mu \,\,\, ,
\eeqa
and\footnote{We do not spell out the details of how to construct the
heavy nucleon EFT from its relativistic counterpart but refer the
reader to the extensive review~\cite{bkmrev}.} 
\beq
\Psi (x) = \exp\{ i mv \cdot x \} \, ( N(x) + h(x) )~.
\eeq
Furthermore, $v_\mu$ denotes the nucleons' four--velocity, 
$S_\mu$ the covariant spin--vector \`a la Pauli--Lubanski
and $g_A$ the axial--vector coupling  constant. We also need 
the generalized pion covariant derivative containing the external
vector ($v_\mu$) and axial--vector ($a_\mu$)  sources,
\beq\label{CDpi}
{\rm d}_\mu U = \partial_\mu U - i(v_\mu+a_\mu+QA_\mu)U
 +iU(v_\mu-a_\mu+QA_\mu) \,\, . 
\eeq
These virtual photon effects can only come in via loop diagrams since
by definition a virtual photon can not be an asymptotic state. Note that
soft photon radiation can be treated by standard methods and will not
be discussed any further.

At second order ($i=2)$, local contact terms with finite low--energy 
constants (LECs) appear.
Following SM, we call these LECs $f_i$ for the heavy baryon approach
and $f_i '$ in the relativistic Lagrangian. As stated before, the
em Lagrangian is given entirely in terms of squares of $Q_{\pm}$ (and
their traceless companions),
\beq\label{L2}
{\cal L}_{\pi N, {\rm em}}^{(2)} 
= \sum_{i=1}^3 \, F_\pi^2 \, f_i ' \,  \bar{\Psi}\, {\cal O}_i^{(2)} \,\Psi
= \sum_{i=1}^3 \, F_\pi^2 \, f_i \,  \bar{N} \,{\cal O}_i^{(2)}\, N~,
\eeq
with the ${\cal O}_i^{(2)}$ monomials of dimension two,
\beq
{\cal O}_1^{(2)}  = \langle \qp^2 - \qm^2 \rangle~, \quad
{\cal O}_2^{(2)}  = \langle Q_+ \rangle \, \qp~, \quad
{\cal O}_3^{(2)}  =  \langle \qp^2 + \qm^2 \rangle~. 
\eeq
Note that we have one term less than given in SM. The third and
fourth term in SM both lead to an overall (unobservable)  em shift $e^2
\bar{\Psi}\Psi$ or $e^2 \bar{N}N$ of the nucleon mass in the
chiral limit and thus can be lumped in one
term. Notice furthermore that only the second term in eq.(\ref{L2}) has an
isovector piece and contributes to the neutron--proton mass
difference~\cite{ms}. The factor $F_\pi^2$ in eq.(\ref{L2}) ensures
that the em LECs have the same dimension as the corresponding strong
dimension two LECs. From the third order calculation of the
proton--neutron mass difference~\cite{ms} one deduces the value
for $f_2$, $f_2 = -(0.45\pm 0.19)\,$GeV$^{-1}$.

The Lagrangian to third order takes the form
\beq \label{L3em}
{\cal L}^{(3)}_{\pi N,{\rm em}} 
= \sum_{i=1}^{12} F_\pi^2 \, g_i' \, \bar \Psi \,{\cal O}_i^{(3)} \, \Psi \,\,  
= \sum_{i=1}^{12} F_\pi^2  \,g_i \, \bar N \,{\cal O}_i^{(3)} \, N \,\, , 
\eeq
with the ${\cal O}_i^{(3)}$ monomials in the fields of dimension three,
tabulated in table~\ref{tab:3}, in their 
relativistic form and the heavy baryon counterparts. Again, for the
$g_i$ to have the same mass dimension as the $d_i$ of the strong sector
defined in ref.\cite{FMS}, we have multiplied them with a factor of
$F_\pi^2$. Thus the $g_i \, (g_i')$ scales as [mass$^{-2}$].
 
\begin{table}[h]
$$
\begin{tabular}{|r|c||c|c|} \hline
i  & Relat.   &     HBCHPT & $\kappa_i$ \\ 
\hline
1  & $  \gamma^\mu \gamma_5  < \qp u_\mu > \, \qp $ & $ 
 < \qp S \cdot u  > \, \qp $ & $4Zg_A(1-g_A^2)$\\
2  & $  \gamma^\mu \gamma_5  < \qm u_\mu > \, \qm $ & $ 
 < \qm S \cdot u  > \, \qm $ & $2g_A(1-g_A^2)(1-2Z)$\\
3  & $  \gamma^\mu \gamma_5  < \qp u_\mu > \, < Q_+ > $ & $ 
 < \qp S \cdot u  > \, < Q_+ > $ &  $-2 g_A$\\
4  & $  \gamma^\mu \gamma_5  < \qp^2 + \qm^2 > \,  u_\mu $ & $ 
 < \qp^2 + \qm^2 > \,  S \cdot u $ & $-g_A (-4-g_A^2)/2$ \\
5  & $  \gamma^\mu \gamma_5  < \qp^2 - \qm^2 > \,  u_\mu $ & $ 
 < \qp^2 - \qm^2 > \,  S \cdot u $ & $g_A (8Z-g_A^2) /2 $\\
6  & $ \gamma^\mu < \qp u_\mu >  \, \qm  $ & $ < \qp v \cdot u > \, \qm $ 
& $-4$ \\
7  & $ \gamma^\mu < \qm u_\mu >  \, \qp  $ & $ < \qm v \cdot u > \, \qp $ 
& $1-3g_A^2$ \\
8  & $ \gamma^\mu < \qm u_\mu >  \, < Q_+ >  $ & 
$ < \qm v \cdot u > \, < Q_+ >  $ & $0$ \\
9 & $ \gamma^\mu \gamma_5 \, [ \qp, [i \na_\mu , \qp ]]  $ &
 $  [ \qp, [i S \cdot \na , \qp ]] $ & $-2$ \\
10 & $ \gamma^\mu \gamma_5 \, [ \qm, [i \na_\mu , \qm ]]  $ &
 $  [ \qm , [i S \cdot \na , \qm ]]  $ & $(1-9g_A^2)/2$ \\
11 & $ \gamma^\mu  \, [ \qp, [i \na_\mu , \qm ]]  $ &
 $  [ \qp , [i v \cdot \na , \qm ]]  $ & $0$ \\
12 & $ \gamma^\mu  \, [ \qm, [i \na_\mu , \qp ]]  $ &
 $  [\qm, [i v \cdot \na , \qp ]]  $ & $0$ \\
\hline
\end{tabular}
$$
\caption{Monomials ${\cal O}_i$ of third order contributing to the em Lagrangian
for the relativistic and heavy baryon formulations. For HBCHPT, the
$\beta$---functions $\kappa_i$  are also given. Here, $Z=C/F_\pi^4$.\label{tab:3}}
\end{table}
Note that we have fewer terms than given in ref.\cite{ms}. First,
we have used the nucleon equations of motion and also dropped all
terms with a single $\langle Q_+\rangle^2$ since these can be lumped together
with the similar terms $\sim \langle \tilde{Q}_+^2 +
\tilde{Q}_-^2\rangle$. For example,
the terms 4 and 6 of table~1 in \cite{ms} both lead to an electromagnetic
renormalization of the axial--vector coupling constant and can thus effectively
be represented by one term. 
The low--energy constants $g_i' (g_i)$ absorb the divergences in the standard
manner (given here for HBCHPT),
\beqa \label{L3LECs}
g_i &=& F_\pi^ 2 \kappa_i \, L + g_i^r (\mu ) \, \, , \\\label{Ldef}
L &=& {\mu^{d-4}\over 16\pi^2} \biggl\{ {1 \over d-4} - {1\over 2}
 \biggl[ \ln(4\pi ) + \Gamma '(1) +1 \biggr] \biggr\} \,\, ,
\eeqa
with $\mu$ the scale of dimensional regularization and $d$ the number
of space--time dimensions. The $g_i^r (\mu)$ are the renormalized,
finite, 
scale--dependent and dimensionless
 low--energy constants. These can  be fixed by data or
have to be estimated with the help of some model. They obey the
standard renormalization group equation.
The $\beta$--functions $\kappa_i$ given in table~\ref{tab:3} agree
with the ones of~\cite{ms} after appropriate rearrangement.
In addition to these terms, there are three which are only needed for
renormalization, i.e. in HBCHPT they are proportional to $\omega =
v\cdot l$, with $l$ the small residual nucleon momentum. The
corresponding monomials and their $\beta$--functions  read
\beqa
{\cal O}_{13}^{(3)} &=& \langle \tilde{Q}_+^2 - \tilde{Q}_-^2 \rangle i
v\cdot \nabla + {\rm h.c.}~, \quad \kappa_{13} = -1/2 -3g_A^2(1+8Z)/4~,
\nonumber \\
{\cal O}_{14}^{(3)} &=& \langle \tilde{Q}_+^2 + \tilde{Q}_-^2 \rangle i
v\cdot \nabla + {\rm h.c.}~, \quad \kappa_{14} = -1/2 + 3 g_A^2/4~, 
\nonumber \\
{\cal O}_{15}^{(3)} &=& \langle {Q}_+ \rangle \tilde{Q}_+ i
v\cdot \nabla + {\rm h.c.}~, \,\,\quad \quad \kappa_{15} = -2~.
\eeqa
In the relativistic formulation, these terms would formally count as
second order.
Note also that one could use the following relations to transfrom some of the 
terms in table~\ref{tab:3}, but this is equivalent to a redefinition of 
a few counterterms and does not reduce the number of terms \cite{ms},
\beqa\label{kurz} 
[\nabla_\mu , Q_\pm] &=& -{i \over 2} [u_\mu , Q_\mp ] + c_\mu^\pm \,\,\, 
\nonumber \\
c_\mu^\pm &=& {1\over 2} \biggl\{ u (\partial_\mu Q - i [v_\mu-a_\mu ,
Q]) u^\dagger \pm u^\dagger (\partial_\mu Q - i [v_\mu +a_\mu ,Q])u 
\biggr\} \,\,\,\, , \\
{}[\nabla^\mu,u_\mu] &=& \frac{i}{2} \, \chi_- - \frac{i}{4} \,
\langle\chi_- \rangle + i\frac{4C}{F^2}\, [Q_+,Q_-]
+ {\cal O}(q^4) \,\, .
\eeqa
where the LEC $C$ is defined in app.~\ref{app:LM}.

\subsection{Fourth  order terms}
\label{sec:L4}

\begin{table}[htb]
$$
\begin{tabular}{|r|c|c|} \hline
i  &  Relat.                    &     HBCHPT \\ \hline
12  & $ \langle \qp^2 + \qm^2 \rangle \, \langle u_\mu u_\nu \rangle \{\na^\mu , \na^\nu \} $ \hc  & 
 $ \langle \qp^2 + \qm^2 \rangle \, \langle (v \cdot u)^2 \rangle $    \\
13 & $ \langle \qp^2 - \qm^2 \rangle \,  \langle u_\mu u_\nu \rangle \{ \na^\mu , \na^\nu \} $ \hc & 
 $ \langle \qp^2 - \qm^2 \rangle \,  \langle (v \cdot u)^2 \rangle  $    \\
14 & $ \langle \qp u_\mu \rangle \, \langle \qp u_\nu \rangle \{\na^\mu , \na^\nu \} $ \hc   & 
 $ \langle \qp v \cdot u \rangle \, \langle \qp v \cdot u \rangle $   \\
15 & $ \langle \qm u_\mu \rangle \, \langle \qm u_\nu \rangle \{\na^\mu , \na^\nu \} $  \hc  &  
 $ \langle \qm v \cdot u \rangle \, \langle \qm v \cdot u \rangle $  \\
16 & $ \langle Q_+  \rangle \, \langle u_\mu u_\nu \rangle \qp \{\na^\mu , \na^\nu \}    $ \hc   & 
$ \langle Q_+  \rangle \, \langle (v \cdot u)^2 \rangle \qp  $    \\
17 & $ \langle Q_+\rangle \, \langle \qp \, u_\mu \rangle \, u_\nu   \{\na^\mu , \na^\nu \}$ \hc & 
$ \langle Q_+\rangle \, \langle \qp \, v \cdot u \rangle \, v \cdot u  $   \\
\hline
18 & $ i \sig  \, \langle \qp [u_\mu , u_\nu ] \rangle \, \qp $  & 
 $ \Sig  \, \langle \qp [u_\mu , u_\nu ] \rangle \, \qp $ \\
19 & $ i \sig  \, \langle \qm [u_\mu , u_\nu ] \rangle \, \qm $  & 
 $ \Sig  \, \langle \qm [u_\mu , u_\nu ] \rangle \, \qm $ \\
20 & $  i \sig  \, \langle \qp^2 + \qm^2 \rangle \, [u_\mu , u_\nu ] $  &
$ \Sig  \, \langle \qp^2 + \qm^2 \rangle \, [u_\mu , u_\nu ] $  \\
21 & $ i \sig  \, \langle \qp^2 - \qm^2 \rangle \, [u_\mu , u_\nu ] $  & 
 $ \Sig  \, \langle \qp^2 - \qm^2 \rangle \, [u_\mu , u_\nu ] $ \\
22 & $ i \sig  \, \langle \qp u_\mu \rangle \, [ \qp , u_\nu ] $  &
$ \Sig  \, \langle \qp u_\mu \rangle \, [ \qp , u_\nu ] $  \\
23 & $ i \sig  \, \langle \qm u_\mu \rangle \, [ \qm , u_\nu ] $  & 
 $ \Sig  \, \langle \qm u_\mu \rangle \, [ \qm , u_\nu ] $ \\
24 & $ i  \sig  \, \langle Q_+ \rangle  \, \langle  \qp \, [ u_\mu , u_\nu ] \rangle $  & 
$ \Sig  \, \langle Q_+ \rangle  \, \langle  \qp \, [ u_\mu , u_\nu ] \rangle $ \\
\hline
25 & $ \gamma^\mu \gamma_5  \, 
\langle \qp \, u_\mu \rangle  \, \langle  \qm \, u_\nu  \rangle i \na^\nu $  \hc  & 
 $  i \, \langle \qp \, S \cdot u \rangle  \, \langle  \qm \,v \cdot u  \rangle $  \\
26 & $ \gamma^\nu \gamma_5  \, \langle \qp \, u_\mu \rangle  \, \langle  \qm \, u_\nu  \rangle 
i \na^\mu $  \hc  & 
 $ i  \, \langle \qp \, v \cdot u \rangle  \, \langle  \qm \, S\cdot u \rangle $   \\
27 & $ \gamma^\mu \gamma_5  \, \langle Q_+ \rangle  \, \langle  \qm \, u_\mu  \rangle  
u_\nu i \na^\nu $  \hc & 
 $  i \, \langle Q_+ \rangle  \, \langle  \qm \, S \cdot u  \rangle  v \cdot u $  \\
28 & $ \gamma^\nu \gamma_5  \, \langle Q_+ \rangle  \, \langle  \qm \, u_\mu  \rangle  u_\nu 
i\na^\mu $  \hc & 
 $ i  \, \langle Q_+ \rangle  \, \langle  \qm \, v \cdot u  \rangle  S \cdot u  $  \\
\hline
\end{tabular}
$$
\caption{``Pion'' terms of fourth order. The terms 18-24 (25-28) involve at least
two (three) pions.\label{tab:strong}}
\end{table}
We now consider the terms of fourth order. We split these into various
classes according to the number and type of external fields.
In the absence of any external fields, we have the so--called ``pion'' terms.
Alltogether, there are
28 such terms from which 11 have the same structure in the relativistic and the
HBCHPT formulations. The corresponding monomials for these particular terms read
\beqa
{\cal O}_1^{(4)} &=&  \langle \qp^2 + \qm^2 \rangle^2 \,\, , \quad
{\cal O}_2^{(4)}  =   \langle  \qp^2 - \qm^2 \rangle^2\,\, , \quad
{\cal O}_3^{(4)}  =   \langle \qp^2 + \qm^2 \rangle\langle 
                                    \qp^2 - \qm^2\rangle \,\, ,
\nonumber \\
{\cal O}_4^{(4)} &=&  \langle \qp^2 + \qm^2 \rangle \langle Q_+ \rangle \, \qp 
\,\, , \quad
{\cal O}_5^{(4)}  =   \langle \qp^2 - \qm^2 \rangle \langle Q_+ \rangle \, \qp 
\,\, , \quad
{\cal O}_6^{(4)}  =   \langle \qp^2 + \qm^2 \rangle\langle u^2 \rangle \,\, ,
\nonumber \\ 
{\cal O}_7^{(4)} &=&   \langle \qp^2 - \qm^2 \rangle\langle u^2 \rangle \,\, ,
\quad
{\cal O}_8^{(4)}  =   \langle \qp u_\mu \rangle\langle \qp u^\mu \rangle \,\, ,
\quad
{\cal O}_9^{(4)}  =   \langle  \qm u_\mu \rangle\langle \qm u^\mu \rangle \,\, ,
\nonumber \\
{\cal O}_{10}^{(4)} &=& \langle Q_+  \rangle\langle u^2 \rangle \qp  \,\, , 
\quad
{\cal O}_{11}^{(4)} = \langle Q_+\rangle \, \langle \qp \, u_\mu \rangle \, 
u^\mu \,\, . 
\eeqa
{}From these, ${\cal O}_{4,5,10}$ are proportional to $\tau^3$, i.e. contribute
differently to reactions involving a proton and a  neutron. Note also that the
first five monomials are all ${\cal O}(e^4)$, they thus involve no small
external momentum or meson mass. While the first term is a pure em shift to
the nucelon mass, the other four operators can contribute to pion--nucleon
scattering through their $\bar{\Psi}\Psi\pi\pi$/$\bar{N}N\pi\pi$--vertices.
The other terms are given in table~\ref{tab:strong}.
We remark that the ``pion'' terms involving the covariant derivatives also
contain contributions from external fields.
Of particular interest are the terms which include explicit chiral
symmetry breaking through the external fields $\chi_+$ and 
$\tilde{\chi}_+$. Note that $\tilde{\chi}_+ = 0$ for $m_u =m_d$. There are
six such terms, which obviously have the same structure in the 
relativistic and the heavy fermion framework. These read:
\beqa
{\cal O}_{29}^{(4)} &=&  \langle \qp^2 + \qm^2 \rangle \tilde{\chi}_+  
\,\, , \quad
{\cal O}_{30}^{(4)}  =   \langle \qp^2 - \qm^2 \rangle  \tilde{\chi}_+
\,\, , \quad
{\cal O}_{31}^{(4)}  =   \langle Q_+ \rangle \, \langle \qp \,   
\tilde{\chi}_+ \rangle  \,\, ,
\\
{\cal O}_{32}^{(4)} &=&  \langle  \qp^2 + \qm^2 \rangle \, \langle 
\chi_+ \rangle  \,\, , \quad
{\cal O}_{33}^{(4)}  =   \langle \qp^2 - \qm^2 \rangle \, \langle
\chi_+ \rangle \,\, , \quad
{\cal O}_{34}^{(4)}  =   \langle Q_+ \rangle \, \qp \,\langle \chi_+ \rangle
\,\, .\nonumber 
\eeqa
Again, only three of these terms are proportional to $\tau^3$, these
are ${\cal O}_{29,30,34}^{(4)}$. The first three of these are $\sim
e^2 (m_u-m_d)$ while the others are $\sim e^2 (m_u+m_d)$. In addition,
we find three terms which to leading order are proportional to
external pseudoscalar sources ($\chi_-, \tilde{\chi}_-$),
\beq
{\cal O}_{35}^{(4)}  =   \langle Q_+ \rangle [ i  \qm  , \, \tilde{\chi}_- ] 
\,\, ,
\quad
{\cal O}_{36}^{(4)}  =   \langle [ \qp , \qm ] \, \tilde{\chi}_- \rangle
\,\, ,
\quad
{\cal O}_{37}^{(4)}  =  [ \qp , \qm ] \, \langle \chi_- \rangle
\,\, ,
\eeq
The terms proportional to the field strength tensor $F_{\mu\nu}^\pm$ build from the
external vector and axial--vector fields 
are tabulated in table~\ref{tab:Fmunu}. These are relevant
for the virtual photon contributions to the magnetic moments, em 
and strange nucleon form factors or in pion photo/electroproduction
off nucleons. 
\begin{table}[htb]
$$
\begin{tabular}{|r|c|c|} \hline
k  &  Relat.   &  HBCHPT   \\ \hline
38  & $ \sig \, \langle \qp^2 + \qm^2 \rangle \, \tilde{F}^+_{\mu\nu} $    & 
$  \Sig \, \langle \qp^2 + \qm^2 \rangle \, \tilde{F}^+_{\mu\nu} $     \\
39  & $ \sig \, \langle \qp^2 - \qm^2 \rangle \,   \tilde{F}^+_{\mu\nu} $  & 
$ \Sig \, \langle \qp^2 - \qm^2 \rangle \,   \tilde{F}^+_{\mu\nu} $    \\
40  & $ \sig \, \langle Q_+ \rangle \,  \langle\qp \,   \tilde{F}^+_{\mu\nu} \rangle $  & 
 $ \Sig \, \langle Q_+ \rangle \,  \langle\qp \,   \tilde{F}^+_{\mu\nu} \rangle $    \\
41  & $ \sig \, \langle \qp^2 + \qm^2 \rangle \, \langle F^+_{\mu\nu} \rangle $    & 
 $ \Sig \, \langle \qp^2 + \qm^2 \rangle \, \langle F^+_{\mu\nu} \rangle $     \\
42  & $ \sig \, \langle \qp^2 - \qm^2 \rangle \,  \langle F^+_{\mu\nu} \rangle $  & 
 $ \Sig \, \langle \qp^2 - \qm^2 \rangle \,  \langle F^+_{\mu\nu} \rangle $    \\
43  & $ \sig \, \langle Q_+ \rangle \, \qp \,   \langle F^+_{\mu\nu} \rangle $  & 
$ \Sig \, \langle Q_+ \rangle \, \qp \,   \langle F^+_{\mu\nu} \rangle $    \\
44  & $ \sig \gamma_5 \, \langle  [ \qp , \qm ]  \tilde{F}^+_{\mu\nu} \rangle $    &  
$ \SIG  \, \langle  [ \qp , \qm ]  \tilde{F}^+_{\mu\nu} \rangle $   \\
45  & $ \sig \gamma_5 \,   [ \qp , \qm ]   \langle {F}^+_{\mu\nu} \rangle $    &  
 $ \SIG \,   [ \qp , \qm ]   \langle {F}^+_{\mu\nu} \rangle $   \\
46  & $ \sig \gamma_5 \, \langle Q_+ \rangle \, [ \qm , \tilde{F}^+_{\mu\nu}] $    & 
 $ \SIG  \, \langle Q_+ \rangle \, [ \qm , \tilde{F}^+_{\mu\nu}] $    \\
\hline
47  & $ \sig \, \langle Q_+ \rangle \, \langle \qm \,  \tilde{F}^-_{\mu\nu} \rangle $  & 
$ \Sig \, \langle Q_+ \rangle \, \langle \qm \,  \tilde{F}^-_{\mu\nu} \rangle $    \\
48  & $ \sig \,  \qp  \, \langle \qm \,  \tilde{F}^-_{\mu\nu} \rangle $  & 
$ \Sig \,  \qp  \, \langle \qm \,  \tilde{F}^-_{\mu\nu} \rangle $    \\
49  & $ \sig \,  \qm  \, \langle \qp \,  \tilde{F}^-_{\mu\nu} \rangle $  & 
$  \Sig \,  \qm  \, \langle \qp \,  \tilde{F}^-_{\mu\nu} \rangle $    \\
50  & $ \sig \gamma_5 \, \langle Q_+ \rangle \, [ \qp , \tilde{F}^-_{\mu\nu}] $    &  
 $ \SIG  \, \langle Q_+ \rangle \, [ \qp , \tilde{F}^-_{\mu\nu}] $   \\
\hline
\end{tabular}
$$
\caption{Fourth order terms with one external photon (38-46)
and one external axial source (47-50). \label{tab:Fmunu}}
\end{table}

In total, we have enumerated 50 terms so far. There are another 40
terms, listed in app.~\ref{app:t4}. These are, however, of no direct relevance
since by use of the equations of motion they can all be transformed
into structures $\sim c_\mu^\pm$, cf. eq.(\ref{kurz}), and are thus only
needed for processes with external axial--vector
fields or low--energy manifestations of $Z^0$--exchange.
Such a procedure would, of course, also lead
to some terms already present in the list discussed so far, i.e. the
corresponding LECs would have to be readjusted. 
So the complete and minimal fourth order pion--nucleon Lagrangian with
virtual photons is given by
\beqa \label{L4em}
{\cal L}^{(4)}_{\pi N,{\rm em}} 
&=& \sum_{i=1}^{5} F_\pi^4 \, h_i' \, 
\bar \Psi \,{\cal O}_i^{(e^4)} \, \Psi \,\,  
+ \sum_{i=6}^{90}  F_\pi^2 \, h_i' \, 
\bar \Psi \,{\cal O}_i^{(e^2 p^2)} \, \Psi \,\,  \no \\ 
&=& \sum_{i=1}^{5} F_\pi^4 \, h_i \, 
\bar \Psi \,{\cal O}_i^{(e^4)} \, \Psi \,\,  
+ \sum_{i=6}^{90}  F_\pi^2 \, h_i \, 
\bar \Psi \,{\cal O}_i^{(e^2 p^2)} \, \Psi \,\,
\eeqa
with the ${\cal O}_i^{(4)}$ monomials in the fields of dimension four,
given in this paragraph and in app.~\ref{app:t4}. To be consistent with the
scaling properties of the dimension two and three LECs, the $h_i$ are
multiplied with powers of $F_\pi^2$ such that the 
first five LECs take dimension [mass$^{-3}$] while the others are 
of dimension [mass$^{-1}$]. The corresponding
$\beta$--functions are defined in analogy to eq.(\ref{L3LECs}) and are called 
$\delta_i$,
\beq\label{L4del}
h_i = h_i^r (\lambda ) + F_\pi^2 \delta_i \, L~.
\eeq
Note that the explicit factor of $\fps$ in front of the
$\delta_i$ appearing in eq.(\ref{L4del})
is due to our convention for the LECs used in eq.(\ref{L4em}). Furthermore, the
strong fourth--order $\beta$--functions determined in ref.\cite{MMS}
are also called $\delta_i$. However, due to the different sturcture
of the strong and em operators at fourth order, no confusion can arise.
Before  applying this EFT to some physical processes, we
briefly discuss the numerical values of the em LECs.

\subsection{Dimensional analysis of the LECs}
\label{sec:diman}

Since very little empirical information exists to pin down the
em LECs $f_i$, $g_i$ and $h_i$ appearing at second, third and
fourth order, we have to resort to dimensional analysis to get
some idea about their values. We had already argued that each power
of a charge matrix $Q$ appearing in any monomial is accompanied by
a factor of $F_\pi$ so that the corresponding LECs have the same
mass dimension as their strong (IC) counterparts. Therefore, the
$f_i$, $h_{6\ldots 90}$, $g_i$ and $h_{1\ldots 5}$ scale as 
[mass$^{-1}$], [mass$^{-1}$],  [mass$^{-2}$] and [mass$^{-3}$], in order. 
Furthermore, the factors of $F_\pi$ are
proportional to the natural low energy scale, $\langle 0|A_\mu^a (0)|\pi^b(x)
\rangle = e^{ipx}p_\mu \delta^{ab}F_\pi$. The non--vanishing of the
pion decay constant in the chiral limit is a necessary and sufficient
condition for spontaneous chiral symmetry breaking. The physical origin
of the em LECs is the integration of hard photon loops. Therefore, each
power in $e^2$, as it is the case in QED, is really a power in the
fine structure constant $\alpha =e^2/4\pi$. Thus, since the natural
scale of chiral symmetry breaking is $\Lambda_\chi \sim m_N \sim M_\rho
\sim 1\,$GeV, we can deduce the following estimates on the renormalized 
em LECs  at the typical hadronic scale which we chose as $M_\rho$ (naturalness
conditions)
\beq\label{diman}
f_i  = \frac{\tilde{f}_i}{4\pi}~, \quad
g_i^r (  M_\rho) = \frac{\tilde{g}_i}{4\pi}~, \quad
h_{1\ldots5}^r (  M_\rho) = \frac{\tilde{h}_{1\ldots5}}{(4\pi)^2}~, \quad
h_{6\ldots90}^r ( M_\rho) = \frac{\tilde{h}_{6\ldots90}}{4\pi}~,
\eeq
with the $\tilde{f}_i$, $\tilde{g}_i$ and $\tilde{h}_i$ are numbers
of order one,
\beq
\tilde{f}_i \sim \tilde{g}_i \sim \tilde{h}_i = {\cal O}(1)~.
\eeq
The $f_i$ are finite and scale--independent since loops only start at third order.
Of course, such type of analysis does not allow to fix the signs of
the LECs. For the estimates given in the next section, we choose these
always to be positive. One should also remember that the numbers of
order one appearing in eq.(\ref{diman}) can be sizeably smaller or
larger than one. From the determination of $f_2$ from the the neutron--proton
mass difference to third order in MS we conclude e.g. that $\tilde{f}_2 = -5.65$.
It would be interesting to develop some model
which would allow one to calculate or estimate these LECs (for an
attempt in the meson sector, we refer to ref.\cite{BU}).

\section{Applications}
\label{sec:app}
\def\theequation{\arabic{section}.\arabic{equation}}
\setcounter{equation}{0}

In this section, we work out the virtual photon effects for various
observables. Although we gave the effective Lagrangian in the
relativistic as well as in the heavy fermion framework, we will
restrict ourselves from here on to HBCHPT. We write the nucleon
four--momentum $p_\mu$ as $p_\mu = mv_\mu + l_\mu$ and consider
small residual momenta, $v\cdot l \ll m$. For more details on the
approach we refer to the review~\cite{bkmrev}. A note on our
nomenclature is necessary. In what follows, we will denote the
strong isospin--conversing contributions simply as ``strong'' and
em stands for all isospin--violating terms, may they be of pure photonic
$(\sim e^{2n})$, pure strong $(\sim (m_u - m_d)^m)$ or mixed
$\sim (e^{2k} (m_u -m_d)^p)$ origin (with $k,m,n,p$ integer). 
This has to be kept in mind in the following.

\subsection{Nucleon self--energy and mass shift}
\label{sec:ZmN}
\begin{figure}[hbt]
   \vspace{0.5cm}
   \epsfxsize=10cm
   \centerline{\epsffile{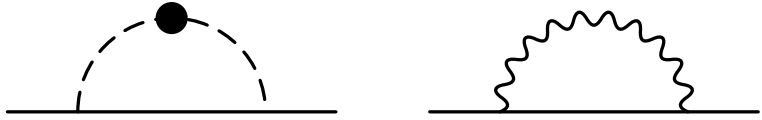}}
   \vspace{1.0cm}
   \centerline{\parbox{15cm}{\caption{\label{fig1}
   One loop graphs contributing to the
   nucleons' electromagnetic self--energy correction. Solid, dashed
   and wiggly lines denote nucleons, pions and virtual photons, in
   order. The heavy dot is an insertion $\sim C$ (iterated to all orders).
  }}}
\end{figure}

\noindent Denoting by $\omega = v\cdot l$ the off--shellness of the nucleon, 
the one--loop third order em contribution to the nucleon self--energy
can be readily evaluated from the two graphs shown in fig.~\ref{fig1}. In
addition, one has to calculate the standard strong contribution (see
e.g. \cite{bkkm,FMSZ}). However, as was already argued in \cite{MMSm},
the natural reference mass in the absence of em effects is the neutral
pion mass, since the pion mass difference is almost entirely of em
origin. Therefore, we express the strong contribution in terms of
the neutral pion mass. In addition, one has to consider the photon loop graph. To
handle the IR divergences related to  the zero photon mass, we
introduce a small photon mass $m_\gamma$ which leads to the
modified photon propagator,
\beq
D_{\mu\nu} (x) = \int \frac{d^dk}{(2\pi)^d} 
\frac{ i g_{\mu\nu}}{k^2 - m_\gamma^2 + i\epsilon}~.
\eeq
Observables are, of course, independent of $m_\gamma$. Putting
pieces together, we find for the self--energy to ${\cal O}(q^3)$
after performing the standard CHPT renormalization, 
\begin{eqnarray}\label{S3}
{\Sigma}^{(3, {\rm str + em})}(\omega)
& = &
\Sigma^{(3, {\rm str + em})}_{\rm loop}(\omega) 
+ \Sigma^{(3, {\rm str + em})}_{\rm div}(\omega)~,
\nonumber \\
{\Sigma}^{(3, {\rm str})}(\omega)
& = &
\frac{3 g_A^2}{32 \pi^2 F_\pi^2} (M_{\pi_0}^2 - \omega^2)
\left[ {\omega} - 2
\sqrt{M_{\pi^0}^2 - \omega^2} \arccos \frac{-\omega}{M_{\pi^0}}
\right] \nonumber\\
&-& \frac{3 g_A^2\omega}{32\pi^2 F_\pi^2}\, (3 M_{\pi^0}^2 - 2 \omega^2)
\ln \frac{M_{\pi^0}}{\lambda}
+ \omega^3 d^r_{24}(\lambda) - 8 M_{\pi^0}^2 \omega d^r_{28}(\lambda)~,\\
{\Sigma}^{(3, {\rm em})}(\omega)
& = &
\frac{g_A^2}{16\pi^2 F_\pi^2} \biggl[ (M_{\pi_+}^2 -M_{\pi_0}^2)
{\omega} - {2}\biggl\{ 
(M_{\pi_+}^2 - \omega^2 )^{3/2} \arccos \frac{-\omega}{M_{\pi^+}} 
\nonumber \\ && \qquad\qquad\qquad\qquad\qquad\qquad\quad - 
(M_{\pi_0}^2 - \omega^2 )^{3/2} \arccos \frac{-\omega}{M_{\pi^0}}
\biggr\}\biggr] \nonumber \\
&-& \frac{g_A^2\omega}{16 \pi^2 F_\pi^2} \biggl[ 
(3M_{\pi^+}^2-2\omega^2) \ln\frac{M_{\pi^+}}{M_{\pi^0}} 
+ 3(M_{\pi_+}^2-M_{\pi_0}^2)\ln\frac{M_{\pi^0}}{\lambda} \biggr]
\nonumber\\
&-& \frac{e^2}{16\pi^2}(1+\tau_3)\biggl
[\omega \biggl( 1 -\ln\frac{m_\gamma}{\lambda}\biggr)
-2\sqrt{m_\gamma^2-\omega^2}\arccos\frac{-\omega}{m_\gamma}\biggr]
\nonumber\\  
&-& e^2 \omega \, \left[ k_{10}^r (\lambda) (1-\tau^3) + 
k_{11}^r (\lambda) + k_{12}^r (\lambda) \right]~.  
\end{eqnarray}
Note for these formulae to be valid, $m_\gamma$ must be larger than
$\omega$ so that for calculating the mass shift $\Sigma (0)$,
one first has to take the limit $\omega \to 0$ before letting the
photon mass vanish.
The result for the strong part agrees with the one in~\cite{FMSZ} (if
one identifies $M_\pi$ in \cite{FMSZ} with $M_{\pi^0}$).
Altogether, a combination of four dimension three (one) em LECs contributes
to the isoscalar (isovector) em self--energy.
The corresponding third order nucleon mass shift follows as
\beqa
\delta m^{(3)} = \Sigma^{(3, \rm str)} (0) + \Sigma^{(3, \rm em)} (0)
&=& -\frac{3g_A^2 M_{\pi^0}^3}{32\pi F_\pi^2}  
-\frac{g_A^2}{16\pi F_\pi^2} \left(M_{\pi^+}^3- M_{\pi^0}^3 \right)
\nonumber \\
&=& -13.6 \qquad - 1.0 \qquad{\rm MeV}~.
\eeqa
To this order, the only em effect comes in via the pion mass
difference, i.e. it is given entirely in terms of the LEC $C$.
Note that it is the same for the  proton and the neutron (as it is
expected from our previous arguments)  and that
it amounts to a 7\% correction to the strong result.

\begin{figure}[htb]
   \vspace{0.5cm}
   \epsfysize=8cm
   \centerline{\epsffile{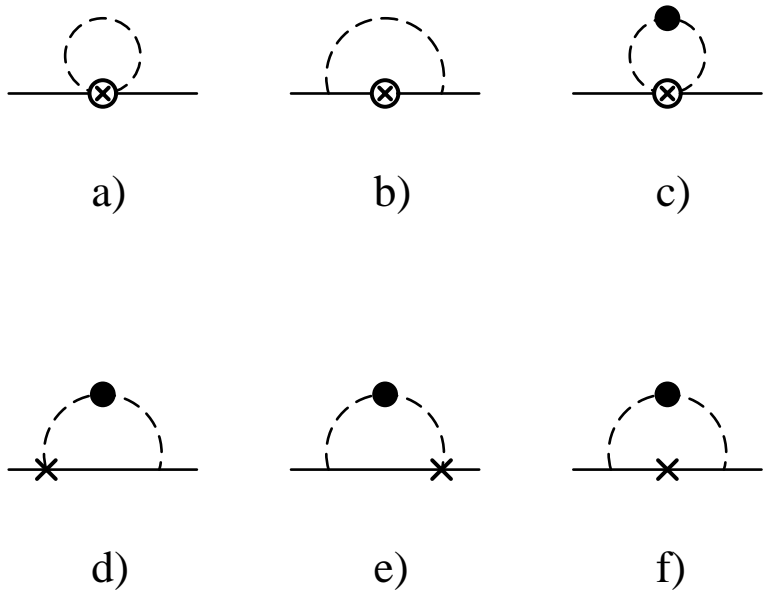}}
   \vspace{0.7cm}
   \centerline{\parbox{15cm}{\caption{\label{fig2}
  Pion loop contributions to the fourth order em (IV) nucleon
   self--energy. The circle--cross (cross) denotes an insertion
   $\sim c_1, c_5, f_1, f_2$ ($\sim 1/m$). For other notations, 
   see fig.1.
  }}}
\end{figure}

\begin{figure}[htb]
   \vspace{0.5cm}
   \epsfysize=8cm
   \centerline{\epsffile{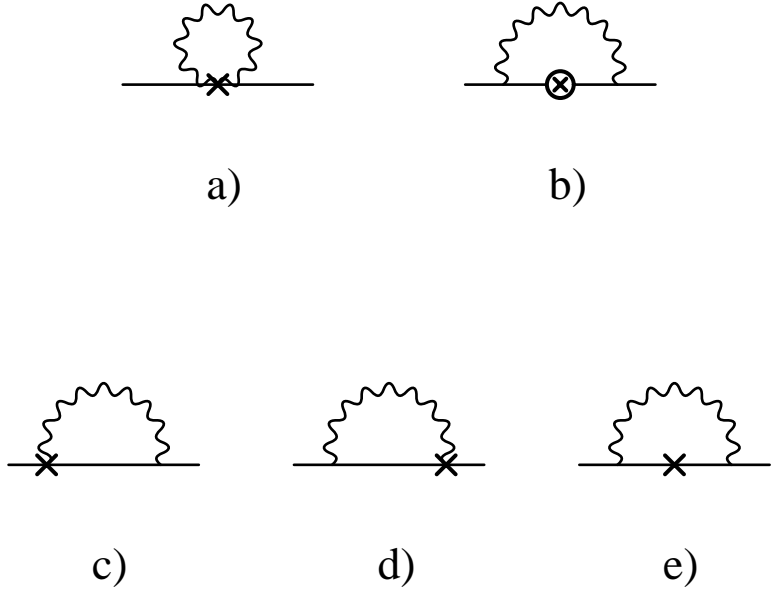}}
   \vspace{0.7cm}
   \centerline{\parbox{15cm}{\caption{\label{fig3}
  Photon loop contributions to the fourth order em (IV) nucleon
  self--energy. For notations,  see figs.1,2.
  }}}
\end{figure}

The fourth order self--energy can be worked out along similar lines.
The much more tedious calculation is detailed in
app.~\ref{app:selfmass}. Besides the pion loop graphs with unequal
masses (i.e. insertions $\sim f_2, c_5, C$),
shown in fig.\ref{fig2}, one has to consider 
photon loop graphs of the tadpole, self--energy and eye graph type,
see fig.\ref{fig3}. These terms introduce additional photonic
counterterms proportional to $m_\gamma^2$, which have to be constructed 
according to symmetry arguments and only drop out at the end.
The corresponding LECs are denoted by $h_{i,\gamma}$ and the pertinent
$\beta$--functions $\delta_{i,\gamma}$.
In addition, there are the so--called induced
terms from lower orders, which chiefly appear because of the
expansion of $\omega$ in powers of $1/m$.  The renormalization of the
self--energy involves combinations of the monomials 
${\cal O}_{29,\ldots,34}^{(4)}$
as well as seven equation of motion terms. These and the
additional photonic monomials of fourth order are
\beqa
{\cal O}^{(4)}_{91} &=&  \stackrel{\leftarrow}{\nabla}_\mu 
\langle Q_+^2 - Q_-^2 \rangle \nabla^\mu \,\, , \quad
{\cal O}^{(4)}_{92}  =   \stackrel{\leftarrow}{\nabla}_\mu 
\langle Q_+^2 + Q_-^2 \rangle \nabla^\mu \,\, , \quad
{\cal O}^{(4)}_{93}  = v \cdot  \stackrel{\leftarrow}{\nabla} 
\langle Q_+^2 - Q_-^2 \rangle v \cdot \nabla \,\, , \nonumber \\
{\cal O}^{(4)}_{94} &=& v \cdot  \stackrel{\leftarrow}{\nabla} 
\langle Q_+^2 + Q_-^2 \rangle v \cdot \nabla \,\, , \quad
{\cal O}^{(4)}_{95}  =   \stackrel{\leftarrow}{\nabla}_\mu 
\langle Q_+  \rangle \tilde{Q}_+  \nabla^\mu \,\, , \quad
{\cal O}^{(4)}_{96}  =   v \cdot \stackrel{\leftarrow}{\nabla} 
\langle Q_+ \rangle \tilde{Q}_+  v \cdot \nabla \,\, ,  \\
{\cal O}^{(4)}_{1,\gamma} &=& m_\gamma^2 \langle Q_+  \rangle
\tilde{Q}_+  \,\,, \quad
{\cal O}^{(4)}_{2,\gamma} = m_\gamma^2 \langle Q_+^2 + Q_-^2  \rangle
\,\, ,\quad
{\cal O}^{(4)}_{3,\gamma} = m_\gamma^2 \langle Q_+^2 - Q_-^2  \rangle
\,\, , \nonumber
\eeqa
and the pertinent $\beta$-functions $\delta_i$ are given
by\footnote{Here, $\delta_{n+m}$ stands for $\delta_n + \delta_m$.}
\beqa
\delta_{1+2+3} &=& 32 f_1 Z - 12(f_1+f_3)g_A^2 Z -
4(f_1+f_2+f_3) + \frac{4}{m} g_A^2 Z^2  + \frac{6}{m} g_A^2 Z^2
- 16 Z^2 (c_2 - \frac{g_A^2}{8m} + 4 c_3 ) \,\, ,\nonumber\\
\delta_{4+5} &=& 12g_A^2 Z - 4(f_1+f_2+f_3) + 8f_2 Z   \,\, \quad
\delta_{29+30} = 12g_A^2Z - 4c_5\,\, , \quad \delta_{31} = -2c_5\,\, ,
\nonumber \\
\delta_{32+33} &=& -8Zc_3 + 4(2Z-1)c_1 -2 Z \bigl( c_2 - \frac{g_A^2}{8m}
\bigr) + \frac{5g_A^2}{4m} Z + 2c_1 g_A^2 Z - \frac{9}{8}(f_1+f_3) + 2 f_1 \,\,
, \\ \delta_{34} &=& -4c_1 + \frac{1}{2} f_2 (1+ \frac{3}{4} g_A^2) \,\, ,
 \quad
\delta_{91+92} = -\frac{2}{m} + \frac{g_A^2}{m}Z\,\, , \quad 
\delta_{93+94} = \frac{4}{m} -3 f_2 \,\, , 
\nonumber \\ \delta_{95} &=& -\frac{2}{m}\,\, , \quad
\quad \delta_{96} = \frac{4}{m} + 9 (f_1+f_3) - \frac{4g_A^2}{m}Z
\,\,  \,\, , \quad
\delta_{1,\gamma} = -\frac{2}{m}\,\, , \quad \delta_{2+3,\gamma} 
= -\frac{2}{m}\,\, .\nonumber
\eeqa
The size of the finite pieces of these new counterterms follows from
the considerations of section~\ref{sec:diman}, $h^r_j (M_\rho) =
{\cal O}(1)/(4\pi )$ $(j=91,\ldots,96)$.
In addition to these, one needs the term~22 of table~1 in ref.\cite{MMS}
to account for the isospin violation through the light quark mass difference.
The corresponding electromagnetic self--energy is given in 
app.~\ref{app:selfmass}. Here, we concentrate on the em  mass shift to
fourth order, which has the general structure\footnote{We have also
calculated the strong fourth order mass shift and find agreement
with the result of ref.\cite{FMSZ} if one identifies $M_\pi$ with
the neutral pion mass as argued before.}
\beqa
\delta m^{(4, {\rm em})} &=& \Sigma^{(4,{\rm em})} (0) +
\Sigma^{(4, {\rm induced})}\nonumber \\
\Sigma^{(4, {\rm induced})} &=&  - \frac{l^2}{2m} \Sigma^{(3, {\rm em})'}
(0) + \delta m^{(2)} \cdot \Sigma^{(3, {\rm em})'} (0) +
\delta m^{(2, {\rm em})} \cdot \Sigma^{(3, {\rm str})'} (0)~,
\eeqa
using the well--known result for the mass shift to second order,
\beq
\delta m^{(2)} = -4c_1 M_{\pi^0}^2 + \delta m^{(2, {\rm em})}
= -4c_1 M_{\pi^0}^2 + \biggl\{
-2B(m_u-m_d) c_5 \tau^3 + \frac{1}{2}e^2 (f_1+f_3) - \frac{1}{2}e^2 \tau^3
f_2 \biggr\}~.
\eeq
It is important to note that the induced mass shift cancels exactly
the strong and photonic eye graph contribution with insertions
$\sim c_{1,5},f_{1,2,3}$ (for details, see app.~\ref{app:selfmass}).
As already noted in ref.\cite{FMSZ},
the so calculated mass shift contains momentum dependent parts
and terms $\sim \ln (M_{\pi^0}/ \lambda )$. With a proper redefinition
of the LECs,
\beq
h_{i,(\gamma)}^r (\lambda) = 
\frac{\delta_{i,(\gamma )}}{16\pi^2} \biggl[ \bar{h}_{i,(\gamma )} + \ln
\frac{M_{\pi^0}}{\lambda} \biggr]~,
\eeq
one can eliminate all these logarithms and also the $l^2$--dependent
terms vanish. This finally gives
\beqa
\delta m^{(4)} &=& \delta m^{(4, \rm str)} + \delta m^{(4, \rm em)}
\nonumber\\
\delta m^{(4, \rm em)} &=& \frac{1}{32\pi^2F_\pi^2} \biggl(
\frac{c_2}{2} - 2 \frac{g_A^2}{m}\biggr) \bigl(M_{\pi^+}^4 -M_{\pi^0}^4
\bigr) \nonumber\\
&+& \frac{M_{\pi^+}^2}{16\pi^2F_\pi^2} \biggl(8c_1 M_{\pi^0}^2 - 
\frac{g_A^2}{m} M_{\pi^+} - (c_2 + 4 
c_3) M_{\pi^+}^2 + 4 e^2 f_1 + e^2 f_2 \tau^3 \biggr) 
\ln \frac{M_{\pi^+}}{M_{\pi^0}} \nonumber \\
\eeqa
with $\delta m^{(4, \rm str)}$ given in eq.(3.22) of ref.\cite{FMSZ}.
Setting $f_1 = \pm 1/(4\pi)$, the numerical value of the em fourth
order mass shift is tiny, for the proton we find $\delta m^{(4, \rm em)}_{p}
= -0.10 \ldots -0.01\,$MeV and for the neutron $\delta m^{(4, \rm em)}_{n}
= -0.06 \ldots +0.03\,$MeV. This is completely negligible and comparable
to the fourth order mass shift, which was shown to be less than 0.1~MeV
in ref.\cite{FMSZ}.

\subsection{Scalar form factor}
\label{sec:ff}

The scalar form factor of the nucleon is defined via
\beq
\langle N(p') |\,  m_u \bar u u + m_d \bar d d \,| N(p)\rangle
= \bar{u} (p') u(p) \, \sigma (t)~, \quad t = (p'-p)^2~,
\eeq
for a nucleon state $|N(p)\rangle$ of four--momentum $p$. At $t=0$, which
gives the much discussed pion--nucleon $\sigma$--term, one can relate
this matrix element to the so--called strangeness content of the 
nucleon. A direct determination of the $\sigma$--term is not possible,
but rather one extends pion--nucleon scattering data into the unphysical
region and determines $\sigma_{\pi N} (t=2M_\pi^2)$, i.e. at the so--called
Cheng--Dashen point. The relation to the $\sigma$--term is given by 
the low--energy theorem of Brown, Peccei and Pardee~\cite{BPP},
\beq
\sigma_{\pi N}(2M_\pi^2) = \sigma_{\pi N}(0) + \Delta \sigma_{\pi N}
+ \Delta R
\eeq
where $\Delta \sigma_{\pi N}$ parametrizes the $t$--dependence of the
sigma--term whereas $\Delta R$ is a remainder not fixed by chiral symmetry.
The most systematic determination of $\Delta \sigma_{\pi N}$ has been
given in ref.\cite{GLS} (although there has been much debate recently
about the pion--nucleon data basis and alike), $\Delta \sigma_{\pi N}
= (15\pm 1)$~MeV. The remainder $\Delta R$ has been bounded in ref.\cite{BKMcd},
$\Delta R < 2$~MeV. In the case of isospin violation, one has of course
to differentiate between the proton and the neutron $\sigma$--terms, as
detailed in ref.\cite{ms}. There, it was shown that the third order
effects can shift the proton $\sigma$--term by about 8\% and have a smaller
influence on the shift to the Cheng--Dashen (CD) point. Here, we work out
explicitly the isospin violating corrections to this shift to fourth
order. This is motivated by the fact that in the difference most of the counterterm
contributions drop out, more precisely, only momentum--dependent contact terms
can contribute to the shift. Such terms only appear at fourth order since
due to parity one needs two derivatives and any quark mass or em insertion
accounts for at least two orders. Before proceeding, we have to discuss
a subtlety related to the definition of the CD--point. Generally, it is
taken to be at $t=2\pps$. For the reasons given before, we choose to
work with $t=2\pns$ to define this point in the unphysical plane. In what
follows, this should be kept in mind (note also that the small effect between
these two values of squared momentum transfer can readily be evaluated).

\begin{figure}[hbt]
   \vspace{0.5cm}
   \epsfxsize=10cm
   \centerline{\epsffile{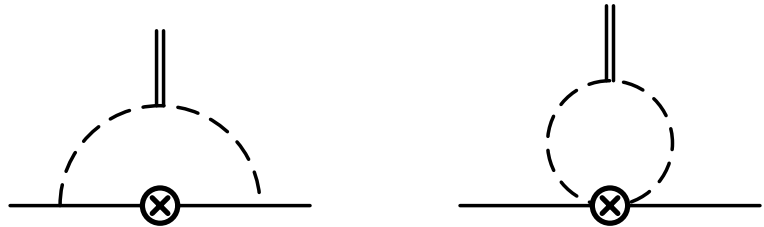}}
   \vspace{1.0cm}
   \centerline{\parbox{15cm}{\caption{\label{fig4}
   One loop graphs contributing to the 
   electromagnetic (IV) self--energy correction for the
   scalar form factor of the nucleon.
   The circle--cross denotes  an insertion $\sim c_5, f_2$.
   For other notations, see fig.1.
  }}}
\end{figure}

We now evaluate the scalar form factor for the proton ($\tau_3 = 1)$.
It can be decomposed as
\beq
\sigma_{\pi N}^{(4)} (t) =  \sigma_{\pi N}^{\rm (4), IC} (t) +
\sigma_{\pi N}^{\rm (4), IV} (t)~.
\eeq
The isospin--conserving strong terms have already been evaluated in
ref.\cite{BKMcd}. Here, we concentrate on the em corrections $\sim
1$ (in isospin space) and all terms $\sim \tau_3$.
Consider first the IV terms. The pertinent one loop graphs shown in fig.\ref{fig4}.
We have eye graphs and tadpoles with insertions $\sim f_2, c_5$. These
can be evaluated straightforwardly and give the terms which
explicitly depend on $\tau_3$,
\beqa
\sigma_{\pi N}^{\rm eye, IV} (t) &=& \left( 2\, B(m_u-m_d)\, c_5
- \frac{1}{2} \,e^2 \,f_2\fps\right) \frac{g_A^2}{4\fps} \, \nonumber \\
&\times& \left[ 2\pps \left(\Omega_3^+ (t) + t\, (\Omega_4^+ (t) 
+ \Omega_2^+ (t) )\right) - \pns \left(\Omega_3^+ (t) + t\, 
(\Omega_4^0 (t) +\Omega_2^+ (t))\right) \right] \, \tau_3
\nonumber \\
\sigma_{\pi N}^{\rm tad, IV} (t) &=& -\left[ B(m_u-m_d)\, c_5 \, I_0^0
  (t)\,\frac{\pns}{\fps}
+\frac{1}{2} \,e^2 \,f_2\, I_0^+ (t) \,{\pps} \right]\, \tau_3~, 
\eeqa
in terms of standard loop functions with the appropriate pion masses
in the propagators as indicated by the superscripts. The momentum dependence
of these equations can be expressed entirely in terms of
\beqa\label{Lfcts}
I_0 (2M_\pi^2) -I_0 (0) &=& -\frac{1}{8\pi^2}\left\{ \frac{\pi}{4}-1\right\}
= 0.0027~,\\
-\frac{t}{4} \Omega_2 (t)\biggl|_{t=2M_\pi^2} 
&=& \frac{\sqrt{2}M_\pi}{128\pi} \ln \frac{2-\sqrt{2}}{2+\sqrt{2}}
= -0.0062 \, M_\pi~,
\eeqa
evaluated for the appropriate pion mass appearing in the loop functions.
 There is no corresponding
tadpole with a photon loop since there is no vertex with two photons and
a scalar source. Furthermore, the photonic eye graph with a $c_5$--insertion
only gives a constant contribution, i.e. it drops out in the scalar form factor.
As counterterm contributions we have terms of order $p^4$, of order
$e^2p^2$ and form ${\cal O}(e^4)$ which contribute to the $\sigma$-term. But 
here we only need the $t$-dependent terms and so we are left with one new LEC,
\beq\label{sigct}
\Delta \sigma_{\pi N}^{\rm ct, IV} =  e^2 \,  M_\pi^2  \, h_{95}^r(\lambda)
   \tau_3~.
\eeq
There is also a strong fourth order counterterm
 $\sim \stackrel{\leftarrow}{\nabla}_\mu  
\tilde{\chi}_+  \nabla^\mu $ , which has a finite low energy constant $d$
and  therefore does not appear in table~1 of
ref.\cite{MMS}. Its effect, however, can be completely absorbed in a
redefinition of the dimension two LEC $c_5$,
\beq
c_5 \to c_5 + \frac{4M_\pi^2\, d}{(4\pi F_\pi)^2}
\eeq
and is thus unobservable. In what follows, we always imply that any
contribution from the strong term $\sim d$ is absorbed in $c_5$ (a similar
remark holds for the term $\sim d_{22}$, see the next section).
Consider now the corresponding em terms $\sim
f_{1,3}$ (i.e. the same loop graphs as in fig.\ref{fig4}) which give
a contribution $\sim 1$ and the pertinent counterterms. These give
\beqa
\sigma_{\pi N}^{\rm (4), IC} (t) &=& 
\frac{1}{2} \,e^2 \,(f_1+f_3) \frac{g_A^2}{4} 
\left[ 2\pps \left(\Omega_3^+ (t)  + t\, (\Omega_4^+ (t) 
+ \Omega_2^+ (t) )\right) \right. \nonumber \\
&& \left. + \pns \left(\Omega_3^+ (t) + t\, 
(\Omega_4^0 (t) +\Omega_2^+ (t))\right) \right] 
- {2} \,e^2 \,f_1\, I_0^+ (t) \,{\pps} \nonumber\\
&&+ \frac{1}{4}e^4 (h_1^r + h_2^r +h_3^r) \fps + 2e^2 \pns
(h^r_{32} + h^r_{33}) + \frac{1}{2} e^2 t  (h_{91}^r +h_{93}^r)  ~.
\eeqa
We now turn towards the numerical analysis of these formulae. 
Consider first the IV terms. Because
of the tiny coefficients appearing in the evaluation of the loop
contributions, cf. eq.(\ref{Lfcts}), these are only fractions of an
MeV, $\Delta \sigma_{\pi N}^{4, IV, \rm loop} = -0.05\,$MeV and can thus
be completely neglected. For the counterterm contributions, setting
all appearing LECs on the values obtained from dimensional
analysis as explained in section~\ref{sec:diman}, one
finds a total contribution $\Delta \sigma_{\pi N}^{4, IV, \rm ct} =
\pm0.01\,$MeV. For the IC em terms, we find
(setting again $f_{1,3} = \pm 1/4\pi$) a completely negligible loop
contribution (less than 0.01~MeV) and the counterterms give 
$\pm 0.7\,$MeV for the LECs estimated from dimensional analysis.
Note, however, that if the numerical factors $\tilde{f}_{1,3}$ are
somewhat bigger than one, one could easily have a shift of $\pm
2\,$MeV, which is a substantial electromagnetic effect.

\subsection{Neutral pion scattering off nucleons}
\label{sec:pi0N}

\begin{figure}[hbt]
   \vspace{0.5cm}
   \epsfxsize=13cm
   \centerline{\epsffile{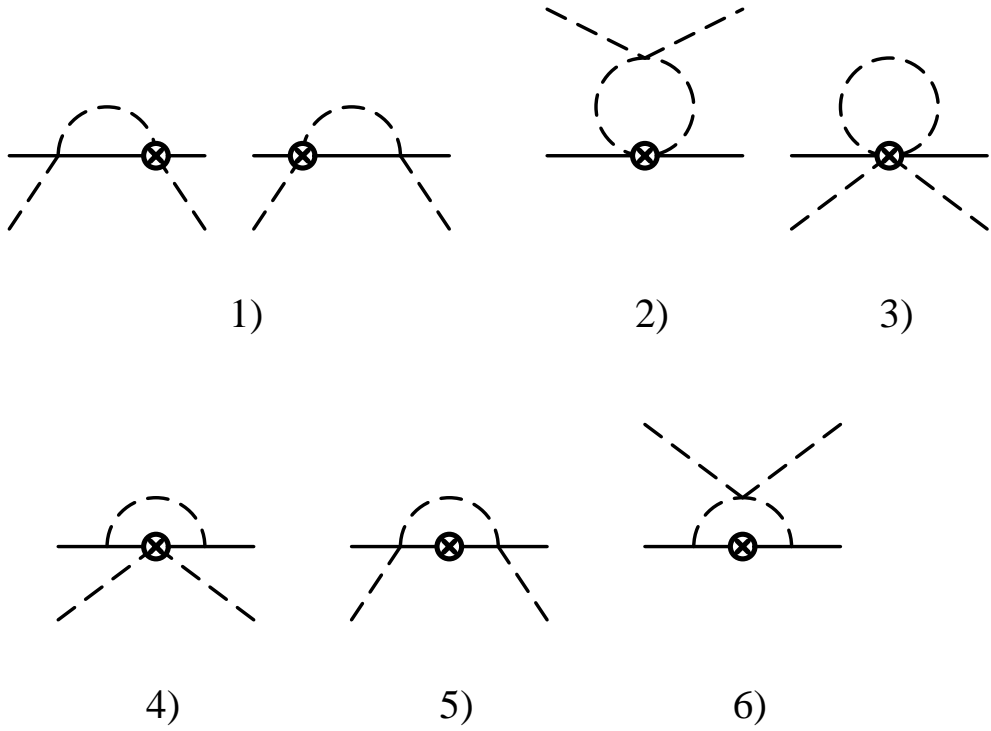}}
   \vspace{1.0cm}
   \centerline{\parbox{15cm}{\caption{\label{visw}
   One loop graphs contributing to the 
   fourth order correction to the scattering lengths
   difference $a(\pi^0 p)-a(\pi^0 n)$. For notations
   see fig.1.
  }}}
\end{figure}

As pointed out long time ago by Weinberg~\cite{weinmass}, the difference
in the S--wave scattering lengths for neutral pions off nucleons is 
sensitive to the light quark mass difference,
\beqa\label{weinf}
a(\pi^0 p)-a(\pi^0 n) &=& \frac{1}{4\pi}\frac{1}{1+M_{\pi^+}/m_p} \,
\frac{-4B(m_u-m_d)c_5}{F_\pi^2} +{\cal O}(q^3) \nonumber \\
&=& \frac{1}{4\pi}\frac{1}{1+M_{\pi^+}/m_p} \,
\Delta_2 (\pn)  +{\cal O}(q^3)~.
\eeqa
It was shown in ref.\cite{ms} by an explicit calculation that to third order
there are no corrections to this formula. This is based on the fact that
the electromagnetic Lagrangian can not contribute at this order
since the charge matrix has to appear quadratic and never two
additional pions can appear. However, at next order one can of course have
loop graphs with one dimension two insertion and additional em
counterterms. To obtain the first correction to Weinberg's prediction,
eq.(\ref{weinf}), one thus has to compute the fourth order 
corrections as it is done here.\footnote{As in
  refs.\cite{weinmass,ms}, we neglect photon
loops and soft photon radiation off the protons. These would affect the Weinberg
prediction and our result in a similar fashion.}
These are due to strong dimension two insertions $\sim c_5$ and em insertions
$\sim f_2$ as shown in fig.\ref{visw}. For the difference $a(\pi^0 p)-a(\pi^0 n)$
we only have to consider the operators $\sim \tau^3$. While all six type of 
graphs contribute
to the strong isospin violation, the graphs 2) and 4) do not have a contribution
$\sim f_2$ because this operator does not couple to neutral pions. 
Expressed in terms of standard loop functions,
we find (the counterterm contribution is discussed below)
\beqa
\Delta_4 (\omega)  &=&  \Delta_2 (\omega ) \, \left( 
1 + \Delta_2^{\rm str} (\omega ) + R \Delta_2^{\rm em} (\omega ) 
\right) + {\cal O}(q^5)~, \\
\Delta_2^{\rm str}(\omega) &=& 
\frac{1}{2F_\pi^2} \left[ \left( J_1^+ (\omega ) + J_1^+ (-\omega)
\right) + \omega \left( J_0^+ (\omega ) - J_0^+ (-\omega) \right) \right]
\nonumber \\
&+& \frac{1}{2F_\pi^2} \left[ M_{\pi^0}^2 I_0^0 -8I_2^0 + \Delta_\pi^+ + 3\Delta_\pi^0
\right] \nonumber \\
&-& \frac{1}{8F_\pi^2} \left[ \omega^2 \left( G_0^+ (\omega) + G_0^+(-\omega)\right)
+2\omega \left(G_2^+ (\omega) + G_2^+(-\omega)\right) + \left(
G_3^+ (\omega) + G_3^+(-\omega)\right)\right]\nonumber \\
&+& \frac{3g_A^2}{4F_\pi^2} \left[ 3\tilde{\Gamma}_2^0 (0) - 2\tilde{\Gamma}_2^+ (0)
- 2M_{\pi^+}^2 \Omega_3^+ (0) -  M_{\pi^0}^2 \Omega_3^0 (0) + 2G_2^0 (0) \right] \\
\Delta_2^{\rm em} (\omega) &=& 
\frac{1}{2F_\pi^2} \left[ \left( J_1^+ (\omega ) + J_1^+ (-\omega)
\right) + \omega \left( J_0^+ (\omega ) - J_0^+ (-\omega) \right) 
+ 2\Delta_\pi^+ \right]
\nonumber \\
&+& \frac{1}{8F_\pi^2} \left[ \omega^2 \left( G_0^+ (\omega) + G_0^+(-\omega)\right)
+2\omega \left(G_2^+ (\omega) + G_2^+(-\omega)\right) + \left(
G_3^+ (\omega) + G_3^+(-\omega)\right)\right]\nonumber \\
&-& \frac{3g_A^2}{4F_\pi^2} \left[ 2\tilde{\Gamma}_2^0 (0) 
+ 2M_{\pi^+}^2 \Omega_3^+ (0) +  M_{\pi^0}^2 \Omega_3^0 (0) - 2G_2^0 (0) \right]~,\\
R &=&  {e^2 \, f_2}/{\Delta_2}~.
\eeqa
where the superscripts $'+,0'$ on the loop functions refer to the pion mass
with which these have to be evaluated (for the precise definitions,
see the review~\cite{bkmrev}). At threshold $\omega = M_{\pi^0}$,
these functions can readily be evaluated. Adding also the pertinent
counterterms, one gets
\beqa
\Delta_2^{\rm str}  &=& \frac{1}{16\pi^2F_\pi^2}\biggl\{ -\pns\llo + \frac{1}{2}
\pps\lm +\frac{3}{2}\pps\llo \nonumber\\
&& \qquad\quad
 -\frac{\pn\pps}{\sqrt{\pps-\pns}}\biggl[ \arccos
\frac{-\pn}{\pp} - \arccos\frac{\pn}{\pp} \biggr] \biggr\}\nonumber \\
&+&\frac{3g_A^2}{16\pi^2F_\pi^2} \biggl\{ -\frac{1}{4}\pps - 3\pns\llo
-\frac{1}{8}\pns \biggr\}~,\\ 
\Delta_2^{\rm em}  &=& \frac{1}{16\pi^2F_\pi^2}\biggl\{ \pns \biggl(
4-15\llo-16\lm\biggr) + \pps\biggl(\frac{5}{2}\llo + \frac{7}{2}\lm\biggr)
\nonumber \\
&& \qquad\quad -\frac{\pn(6\pps - 8\pns )}{\sqrt{\pps-\pns}}\biggl[ \arccos
\frac{-\pn}{\pp} - \arccos\frac{\pn}{\pp} \biggr] \biggr\}\nonumber \\
&+&\frac{3g_A^2}{16\pi^2F_\pi^2} \biggl\{ \pns \biggl(\frac{1}{4} +2\llo
\biggr) +2\pps \biggl(\frac{1}{4} + \llo + \lm \biggr)\biggr\}~, \\ 
\Delta_4^{\rm ct} (\pn)  &=& -\frac{2}{\fps}B(m_u-m_d)\left\{
e^2( h_{29}^r
(\lambda) + h^r_{30}(\lambda) )\right\}~, 
\eeqa
where we have no exhibited the strong fourth order contribution proportional
to $d_{22}$ since it can be absorbed in the $c_5$ term and is thus unobservable
as discussed before.
Consider first the loop contributions. Since we can not fix the counterterms
from data, we are left with a spurious scale dependence which reflects the
theoretical uncertainty at this order. For $\lambda = \{0.5, 0.77, 1.0\}\,$GeV
we find
\beq
\Delta_2^{\rm str} = \{ -7.1 , 0.9 , 5.7 \} \cdot 10^{-2}~, \quad
\Delta_2^{\rm em}  = \{ 11.5 , 12.0 , 12.3 \}\cdot 10^{-2} ~.
\eeq
In both cases, the dominant contribution comes from the rescattering
graphs~5) which is largely canceled the pion rescattering diagrams~6).
The counterterms are estimated based on dimensional analysis at the
scale $\lambda = M_\rho$ and give a contribution of about $-0.3  
\cdot 10^{-2}$. Even if the LECs $ h_{29}^r (\lambda) + h^r_{30}(\lambda)$
would be a factor of ten larger than assumed, the counterterm contribution
would not exceed $\pm 3\%$.
 Altogether, the correction Weinberg's prediction, eq.(\ref{weinf}),
are in the range of $4$ to $18$ percent, i.e. fairly small.   
We do not attempt here to calculate the isoscalar fourth order
corrections  since at present the sizeable uncertainty in the
determination of the isoscalar scattering lengths extracted from level
shifts in pionic hydrogen and deuterium is larger than this
isospin--violating effect (for a detailed discussion, see e.g.
ref.\cite{FMSi}). Finally, we wish to mention that in ref.\cite{AMB}
isospin--violation for neutral pion photoproduction off nucleons
was discussed which allows one to eventually measure directly
the very small $\pi^0 p$ scattering length by use of the final--state
theorem.

\section{Summary and conclusions}
\label{sec:summ}
\def\theequation{\arabic{section}.\arabic{equation}}
\setcounter{equation}{0}

In this paper, we have considered baryon chiral perturbation theory in the
presence of virtual photons to fourth order in small momenta $q$ (counting the
electric charge as a small parameter $e \sim q$). The pertinent results of
this investigation can be summarized as follows:
\begin{enumerate}
\item[i)] We have constructed the complete fourth order electromagnetic
Lagrangian including up to four nucleon (quark) charge matrices. We have
omitted all terms which only lead to an overall mass shift or coupling
constant renormalization. The corresponding terms are collected in 
section~\ref{sec:L4} and in app.~\ref{app:t4}.
\item[ii)] To get an estimate of the novel electromagnetic low--energy constants,
we have performed dimensional analysis and argued that measured in
appropriate powers of the inverse scale of chiral symmetry breaking,
these should be of order $1/4\pi$ or $1/(4\pi)^2$.
\item[iii)] We have evaluated the third order isospin--violating 
nucleon mass shift and found that amounts to a seven percent correction of its 
strong  (isospin--conserving) counterpart. The fourth order electromagnetic
mass shift is tiny (as is the strong fourth order mass shift).
\item[iv)] The electromagnetic isospin--conserving
contributions to the scalar form factor of the proton (and neutron)
are of the order 1$\ldots$2~MeV, i.e. not completely negligible 
and are thus of (minor) relevance in connecting the $\sigma$--term at the
Cheng--Dashen point to its value at zero momentum transfer.
On the other hand, all isospin--violating contributions
are very small, only fractions of an MeV.
\item[v)] We have worked out the first corrections to Weinbergs
time--honored  prediction for the difference of the S--wave scattering
for the neutral pions off nucleons. These corrections are small, we
estimate their numerical value in the range from 4 to 18\%.
\end{enumerate}

\noindent

\bigskip

\noindent {\bf Acknowledgments}

\noindent We are grateful to Evgeni Epelbaum for some useful comments.

\vspace{1cm}

\appendix
\def\theequation{\Alph{section}.\arabic{equation}}
\setcounter{equation}{0}
\section{Electromagnetic meson Lagrangian}
\label{app:LM}

Here we briefly review the effective electromagnetic meson Lagrangian,
following ref.\cite{MMSm}.
The effective field theory build of pions, collected in $U(x) = u^2(x)$,
photons $(A_\mu$) and other scalar
($s$), pseudoscalar ($p$), vector ($v_\mu$) and axial--vector
($a_\mu$) external sources starts at dimension two,
\beq \label{Lm2}
{\cal L}^{(2)}_{\pi\pi} = - {1\over 4}F_{\mu\nu}F^{\mu\nu} - {\lambda \over 2} 
(\partial_\mu A^\mu )^2 + {F^2\over 4} \langle {\rm d}_\mu U {\rm d}^\mu
 U^\dagger + \chi U^\dagger + \chi^\dagger U \rangle + C \langle Q_R U
 Q_L U^\dagger \rangle \,\, , 
\eeq
with $F = 88\,$MeV the pion decay constant in the chiral limit,
$F_{\mu\nu} = \partial_\mu A_\nu - \partial_\nu A_\mu$ denotes the photon field
strength tensor, $\lambda$ is the gauge--fixing parameter (from here on,
 we work in the Landau gauge $\lambda = 1$), d$_\mu$ the generalized covariant
derivative (given in eq.(\ref{CDpi}))
and  $Q$ is  the quark charge matrix, see eq.(\ref{qmes}).
It is advantageous to work  in the so--called $\sigma$--model gauge,
\beq
U(x) = \sigma(x) \, {\bf 1} + i \vec{\tau} \cdot \vec{\pi} (x) /F   
\,\, , \quad \sigma(x) = \sqrt{1 - \pi^2 (x)/F^2 } \,\, .
\eeq
The third term in eq.(\ref{Lm2}) is the standard non--linear
 $\sigma$--model coupled to external sources (we neglect here
singlet components and set $\langle a_\mu \rangle = \langle v_\mu \rangle  
=0$).  The last term in the dimension two
Lagrangian is the lowest order chiral invariant term one can
construct from pion and photon fields \cite{gepdR}. To make it
invariant under chiral SU(2)$_L\times$SU(2)$_R$ transformations, one
has introduced the spurions $Q_{L,R}$ given in eq.(\ref{Spurm}).
The constant $C$ can be calculated from the neutral to charged pion mass 
difference since this term leads to 
\beq
(M_{\pi^+}^2 - M_{\pi^0}^2)_{\rm em} \equiv
 (\delta M^2_{\pi^\pm})_{\rm em} = \frac{2e^2C}{F^2}~, \quad
C = 5.9\cdot 10^{-5}\, {\rm GeV}^4~.
\eeq
This  identification is based upon the fact that the quark mass difference
$m_d - m_u$ only gives a tiny contribution to $M^2_{\pi^+} -
M^2_{\pi^0}$ due $\pi^0-\eta$--mixing. Note that in the 
$\sigma$--model gauge, this is the only contribution from the term $\sim C$. 
It is useful to introduce the dimensionless constant $Z = C / F^4  = 0.89$.
As already stated before, we count
the external vector and axial--vector fields as well as the charge
matrices $Q, Q_L, Q_R$ as ${\cal O}(q)$ and the photon field as
${\cal O}(1)$. This has the advantage of a consistent power counting
between the strong and electromagnetic interactions, i.e. $e \sim q$
and one has terms of dimension two, four and so on. Here, dimension
two means either order $q^2$ or $e^2$ and similar at higher orders.
The fourth order em Lagrangian for SU(2) has been developed in ref.\cite{MMSm}.
It is a sum of 13 local operators quadratic in the charge matrix,
\beq \label{L4}
{\cal L}^{(4)}_{\pi\pi, \rm em} = \sum_{i=1}^{13} k_i \, {\cal O}_i \,\, , 
\eeq
with the ${\cal O}_i$ monomials in the fields of dimension four. The
low--energy constants $k_i$ absorb the divergences in the standard
manner,
\beq \label{L4mem}
k_i = \kappa_i \, L + k_i^r (\mu ) \, \, .
\eeq
with $L$ given in eq.(\ref{Ldef}). The corresponding expressions for the
$O_i$ and the corresponding $\beta$--functions $\kappa_i$ are given
in \cite{MMSm}. More details can be found in ref.\cite{KU}.

\def\theequation{\Alph{section}.\arabic{equation}}
\setcounter{equation}{0}
\section{Self--energy  of the nucleon to fourth order}
\label{app:selfmass}

In this appendix, we give the explicit result for the em contribution
to the nucleons' self--energy to fourth order, which follows after
straightforward but tedious calculation of the diagrams shown in
figs.\ref{fig1},\ref{fig2}, the induced terms from the lower
orders and the pertinent counterterms. The resulting lengthy expression
takes the form (using as always dimensional regularization)
\beq
\Sigma^{(4,em)} (\omega ) = \sum_{i=1}^7 
\Gamma_i \, \tilde{\Sigma}^{(4,em)}_i (\omega )~,
\eeq
with
\beqa
\tilde{\Sigma}^{(4,em)}_1 (\omega ) =  &\biggl\{&
\bigl( l^2 - 2\omega^2 + 8mc_1M_{\pi^0}^2 \bigr) \biggl[
3 \omega \sqrt{M_{\pi^+}^2 - \omega^2} \arccos \frac{-\omega}{M_{\pi^+}}  
-3 \omega \sqrt{M_{\pi^0}^2 - \omega^2} \arccos \frac{-\omega}{M_{\pi^0}}
\biggr]  \nonumber\\  
&-&\frac{1}{2} (M_{\pi^+}^2 - M_{\pi^0}^2 ) 
\biggl[ l^2 -\omega^2 + 8m c_1
M_{\pi^0}^2 + M_{\pi^+}^2 + M_{\pi^0}^2 \biggr] 
\nonumber\\
&+& 3M_{\pi^+}^2 \omega \sqrt{M_{\pi^+}^2 - \omega^2} 
\arccos \frac{-\omega}{M_{\pi^+}}  
- 3M_{\pi^0}^2 \omega \sqrt{M_{\pi^0}^2 - \omega^2} 
\arccos \frac{-\omega}{M_{\pi^0}} \nonumber\\
&-&\frac{3}{2}( M_{\pi^+}^2 - M_{\pi^0}^2 )    
\biggl[ l^2 -4\omega^2 + 8m c_1
M_{\pi^0}^2 + M_{\pi^+}^2 + M_{\pi^0}^2 \biggr]\ln\frac{M_{\pi^0}}{\lambda} 
\nonumber\\
&+& \frac{3}{8} (M_{\pi^+}^4 - M_{\pi^0}^4 ) \ln\frac{M_{\pi^0}}{\lambda}
+\frac{3}{8} M_{\pi^+}^4 \ln\frac{M_{\pi^+}}{M_{\pi^0}} \nonumber\\
&+& \frac{1}{2} \biggl[ l^2 -2\omega^2 + 8m c_1
M_{\pi^0}^2 + M_{\pi^+}^2  \biggr] \bigl( 6\omega^2 -3M_{\pi^+}^2 
\bigr) \ln\frac{M_{\pi^+}}{M_{\pi^0}} \biggr\}~, \\
\tilde{\Sigma}^{(4,em)}_2 (\omega ) =  
&\biggl\{& \biggl[ 8c_1 \pns -
\bigl( c_2 - \frac{g_A^2}{8m} \bigr) \pps - 4c_3 \pps \biggl] \pps
\ln\frac{M_{\pi^+}}{M_{\pi^0}} \nonumber \\
&+& \biggl[ 8c_1 \pns -
\bigl( c_2 + 4c_3 - \frac{g_A^2}{8m} \bigr) (\pps + \pns ) 
\biggl] (\pps - \pns )
\ln\frac{M_{\pi^0}}{\lambda} \nonumber \\
&+& \frac{1}{64\pi^2} (\pp^4 - \pn^4 ) c_2~,\\
\tilde{\Sigma}^{(4,em)}_3 (\omega ) =
&\biggl\{& \frac{1}{2} e^2 (f_1+f_3)  \biggr\}\times \nonumber \\  
&\biggl\{& \frac{1}{2} \bigl(-6\pps -3\pns + 18\omega^2 \bigr)
\ln\frac{M_{\pi^0}}{\lambda} + \frac{1}{2} \bigl(-6\pps +12 \omega^2
\bigr)\ln\frac{M_{\pi^+}}{M_{\pi^0}} \no \\
&&+ 3\omega \biggl[ 2\sqrt{\pps-\omega^2} \arccos \frac{-\omega}{\pp}
+ \sqrt{\pns-\omega^2} \arccos \frac{-\omega}{\pn} \biggr] \no\\
&& - \frac{1}{2} \bigl(2\pps + \pns + 3\omega^2\bigr)\biggr\}~,\\
\tilde{\Sigma}^{(4,em)}_4 (\omega ) =
&\biggl\{&  2 B (m_u -m_d) c_5 + \frac{1}{2} e^2 f_2   \biggr\} 
\times \nonumber \\  
&\biggl\{& \frac{1}{2} \bigl( -3  \pns + 6\pps - 6\omega^2 \bigr)
\ln\frac{M_{\pi^0}}{\lambda} + \frac{1}{2} \bigl( 6 \pps - 12 \omega^2
\bigr)\ln\frac{M_{\pi^+}}{M_{\pi^0}} \no \\
&&+ 3\omega \biggl[ -2\sqrt{\pps-\omega^2} \arccos \frac{-\omega}{\pp}
+ \sqrt{\pns-\omega^2} \arccos \frac{-\omega}{\pn} \biggr] \no\\
&& - \frac{1}{2} \bigl( -2\pps + \pns - \omega^2\bigr)\biggr\}~,\\
\tilde{\Sigma}^{(4,em)}_5 (\omega ) =  
&\biggl[& 4 e^2 f_1 \pps + (2B(m_u -m_d ) c_5 + e^2 f_2 \pps )\tau^3
\biggr] \ln\frac{M_{\pi^0}}{\lambda} \no \\ &+& \biggl[ 4f_1 e^2 \pps  
+ e^2 f_2 \pps \tau^3 \biggr] \ln\frac{M_{\pi^+}}{M_{\pi^0}}  \\
\tilde{\Sigma}^{(4,em)}_6 (\omega ) =  
&\biggl\{& -(-2\omega^2 -l^2 -8c_1 m \pns - m_\gamma^2) \biggl(
\ln\frac{m_\gamma}{\lambda} +1 \biggr) \no\\
&+& \omega \biggl( \frac{1}{\sqrt{m_\gamma^2-\omega^2}} (\omega^2 -l^2
-8c_1 m \pns ) - \sqrt{m_\gamma^2-\omega^2} \biggr) \arccos 
\frac{-\omega}{m_\gamma} \, \biggr\} \\
\tilde{\Sigma}^{(4,em)}_7 (\omega ) =
&\biggl\{& \frac{1}{2} e^2 m_\gamma^2 (h_{2,\gamma}^r (\lambda) +
h_{3,\gamma}^r (\lambda)) +  \frac{1}{2} e^2 m_\gamma^2 
h_{1,\gamma}^r (\lambda ) \tau^3 \no\\
&+& \frac{2}{F_\pi^2} B(m_u -m_d) \bigl[ 4\pns  d_{22}^r
(\lambda) + \omega^2 d_{97}^r (\lambda) \bigr] \tau^3 \no\\
&+&  B(m_u -m_d) e^2 \bigl[2 h_{31}^r (\lambda) +
 (h_{29}^r (\lambda) +  h_{30}^r (\lambda))\tau^3  \bigr] \no \\
&+& \frac{1}{4} e^4 \bigl[ h_{1}^r (\lambda) +  h_{2}^r (\lambda)
+ h_{3}^r (\lambda) +  (h_{4}^r (\lambda)+ h_{5}^r (\lambda) )\tau^3 \bigr] 
\no\\
&+& \frac{1}{2}e^2 \biggl[ 
l^2 \bigl[ h_{91}^r (\lambda)+ h_{92}^r (\lambda)
+ h_{95}^r (\lambda) \tau^3 \bigr] 
+ \omega^2 \bigl[ h_{93}^r (\lambda)+ h_{94}^r (\lambda)
+ h_{96}^r (\lambda) \tau^3 \bigr] \biggr]\no \\
&+& 2e^2 \pns \bigl[ h_{34}^r (\lambda) \tau^3 + 
( h_{32}^r(\lambda) + h_{33}^r (\lambda) )
\bigr] \, \biggr\}~, \\
\eeqa
with
\beq
\Gamma_1 = \frac{g_A^2}{m \Lambda^2_\chi}~, \,\,
\Gamma_2 = \frac{1}{\Lambda_\chi^2}~,\,\,
\Gamma_3 = \frac{g_A^2}{\Lambda_\chi^2}~,\,\, 
\Gamma_4 = \Gamma_5 = \Gamma_2~, \,\,
\Gamma_6 = -\frac{e^2}{m \Lambda_\chi^2}(1+\tau^3)~,\,\, 
\Gamma_7 = -1 \,\, ,
\eeq
and 
\beq 
\Lambda_\chi^2 = 16 \pi^2  F_\pi^2 \quad .
\eeq
The first two contributions $\Sigma_{1,2} (\omega )$
 are due to pion loops with an insertion
$\sim C$, i.e. the charged to neutral pion mass difference. The
next two, $\Sigma_{3,4,5} (\omega )$,
 are pion loops with isospin--violating insertions
$\sim c_5, f_{1,2,3}$. The photon loop graphs  are collected
in $\Sigma_6 (\omega )$ and the last term $\Sigma_7 (\omega )$
is the contribution from the various strong, virtual photon and 
``photonic'' ($\sim m_\gamma^2 )$ counterterms at the scale $\lambda$.

\def\theequation{\Alph{section}.\arabic{equation}}
\setcounter{equation}{0}
\section{Additional fourth order terms}
\label{app:t4}

Here, we enumerate the additional fourth order terms already mentioned
in sec.~\ref{sec:L4}. We give the number of the monomial ($i=51,\ldots,90)$, 
its relativistic and its heavy baryon form,
\beqa
&51\qquad&   \langle [i\na_\mu ,\qp] \, [ \qm ,u^\mu] \rangle~, \quad     
 \langle [i\na_\mu ,\qp] \, [ \qm ,u^\mu] \rangle~, \no \\
&52\qquad&  \langle [i\na_\mu ,\qm] \, [ \qp ,u^\mu] \rangle~, \quad 
 \langle [i\na_\mu ,\qm] \, [ \qp ,u^\mu] \rangle~, \no   \\
&53\qquad& \langle [i\na_\mu ,\qp] \, [ \qm ,u_\nu] \rangle   \{\na^\mu , 
\na^\nu \} \hc~, \quad   
\langle [iv \cdot \na ,\qp] \, [ \qm , v \cdot u] \rangle~, \no   \\
&54\qquad& 
 \langle [i\na_\mu ,\qm] \, [ \qp ,u_\nu] \rangle \{\na^\mu , \na^\nu \} \hc~,
\quad \langle [iv \cdot \na ,\qm] \, [ \qp , v\cdot u] \rangle~, \no\\
&55\qquad& 
\langle [i\na_\mu , u_\nu ] \, [ \qp , \qm] \rangle  \{\na^\mu , \na^\nu \} \hc~,
\quad     \langle [iv \cdot \na , v \cdot u ] \, [ \qp , \qm] \rangle~, \no  \\
&56\qquad&  \langle \qm \,  [\na^\mu , [ \na_\mu , \qm ]] \rangle~, \quad
 \langle \qm \,  [\na^\mu , [ \na_\mu , \qm ]] \rangle~, \no\\
&57\qquad&
 \langle \qm \,  [\na_\mu , [ \na_\nu , \qm ]] \rangle\{\na^\mu , \na^\nu \} \hc~,
\quad \langle \qm \,  [v \cdot \na , [v \cdot  \na , \qm ]] \rangle~, \no \\
&58\qquad& \langle [\na_\mu , \qm ] \,  [ \na^\mu , \qm ]] \rangle~, \quad 
 \langle [\na_\mu , \qm ] \,  [ \na^\mu , \qm ]] \rangle~, \no \\
&59\qquad&  \langle [\na_\mu , \qm ] \,  [ \na_\nu , \qm ]] \rangle\{\na^\mu , \na^\nu
\} \hc~, \quad  \langle [v \cdot \na , \qm ] \,  [ v\cdot \na , \qm ]] \rangle~,\no \\
&60\qquad&  \langle Q_+  \rangle  [ [i \na_\mu, \qm]  , u^\mu ]~, \quad 
 \langle Q_+  \rangle    [ [i \na_\mu, \qm]  , u^\mu ]~,  \no  \\
&61\qquad& \langle Q_+  \rangle  [ [i \na_\mu, \qm]  , u_\nu ] \{\na^\mu , \na^\nu 
\} \hc~,  \quad \langle Q_+  \rangle  [ [iv\cdot  \na , \qm] , v \cdot u ]~,\no \\
&62\qquad& \langle Q_+  \rangle [ [i \na_\mu,  u_\nu ]  , \qm  ] \{\na^\mu ,
\na^\nu \} \hc~, \quad \langle Q_+  \rangle [ [ iv \cdot \na,  v \cdot u ]  
, \qm  ]~, \no \\ 
&63\qquad& \langle  [ \na^\mu,  \qp ] [\na_\mu  , \qp  ] \rangle~, \quad
 \langle  [ \na^\mu,  \qp ] [\na_\mu  , \qp  ] \rangle~, \no  \\
&64\qquad&\langle 
[ \na_\mu,  \qp ] [\na_\nu  , \qp ] \rangle \{\na^\mu , \na^\nu \} \hc~,
\quad  \langle  [ v \cdot \na ,  \qp ] [ v\cdot \na  , \qp  ] \rangle~, \no \\
&65\qquad& \langle  \qp [ \na^\mu ,  [\na_\mu  , \qp  ]] \rangle~, \quad 
\langle  \qp [ \na^\mu ,  [\na_\mu  , \qp  ]] \rangle~, \no   \\
&66\qquad& \langle  \qp [ \na_\mu ,  [\na_\nu  , \qp  ]] \rangle \{\na^\mu , \na^\nu
\} \hc~, \quad  
 \langle  \qp [ v \cdot \na ,  [v \cdot \na  , \qp  ]] \rangle~, \no  \\
&67\qquad& [ \na^\mu , [ \na_\mu , \qp ]]  \, \langle Q_+\rangle~, \quad
 [ \na^\mu , [ \na_\mu , \qp ]]  \, \langle Q_+\rangle~, \no   \\
&68\qquad& 
[ \na_\mu , [ \na_\nu , \qp ]]  \, \langle Q_+\rangle \{\na^\mu , \na^\nu \} \hc~,
\quad  [ v \cdot \na , [ v \cdot \na , \qp ]]  \, \langle Q_+\rangle~,\no    \\
&69\qquad& \sig  \,  [\na_\mu , \qp ] \, \langle \qm u_\nu \rangle~, \quad
 i\Sig  \,  [\na_\mu , \qp ] \, \langle \qm u_\nu \rangle~, \no \\
&70\qquad& \sig  \,  \qp \, \langle   [\na_\mu , \qm ] \, u_\nu \rangle~, \quad 
 i\Sig  \,  \qp \, \langle   [\na_\mu , \qm ] \, u_\nu \rangle~,\no\\
&71\qquad& 
\sig  \,  \langle Q_+ \rangle \, \langle   [\na_\mu , \qm ] \, u_\nu \rangle~,
\quad i\Sig  \,\langle Q_+ \rangle \,\langle [\na_\mu , \qm ] \, u_\nu \rangle~,\no \\
&72\qquad& \sig  \,  [\na_\nu , \qm ] \, \langle \qp u_\nu \rangle~,\quad
 i\Sig  \,  [\na_\nu , \qm ] \, \langle \qp u_\nu \rangle~, \no   \\
&73\qquad& \sig  \, \qm \, \langle   [\na_\mu , \qp ] \, u_\nu \rangle~, \quad
 i\Sig  \,  \qm \, \langle   [\na_\mu , \qp ] \, u_\nu \rangle~, \no \\
&74\qquad& \sig  \,  u_\mu \, \langle   [\na_\nu , \qp ] \, \qm \rangle~, \quad
i\Sig  \,  u_\mu \, \langle   [\na_\nu , \qp ] \, \qm \rangle~, \no \\
&75\qquad& \sig  \,  u_\mu \, \langle   [\na_\nu , \qm ] \, \qp \rangle~, \quad
 i\Sig  \,  u_\mu \, \langle   [\na_\nu , \qm ] \, \qp \rangle~, \no \\
&76\qquad&  i\sig \, [ [ \na_\mu ,\qp] , \,  [\na_\nu , \qp ] ]~, \quad 
 \Sig  \, [ [ \na_\mu ,\qp] , \,  [\na_\nu , \qp ] ]~, \no \\
&77\qquad&  i\sig  \,  [ [ \na_\mu ,\qm] , \,  [\na_\nu , \qm ] ]~, \quad
 \Sig  \,  [ [ \na_\mu ,\qm] , \,  [\na_\nu , \qm ] ]~, \no \\
&78\qquad& i\sig  \,  [ \qp , \,  [ \na_\mu , [\na_\nu , \qp ] ]]~, \quad 
 \Sig  \,  [ \qp , \,  [ \na_\mu , [\na_\nu , \qp ] ]]~, \no \\
&79\qquad&  i\sig  \,  [ \qm , \,  [ \na_\mu , [\na_\nu , \qm ] ]]~, \quad 
  \Sig  \,  [ \qm , \,  [ \na_\mu , [\na_\nu , \qm ] ]]~, \no \\
&80\qquad& 
\gamma^\mu \gamma_5 \, \langle Q_+  \rangle \, [ [\na_\mu , \qp] \, , u_\nu ] 
 \na^\nu \hc~, \quad
\, \langle   Q_+  \rangle  \, [ [S \cdot \na , \qp] \, , v \cdot u  ]~,\no  \\
&81\qquad&\gamma^\nu \gamma_5  \, \langle Q_+ \rangle \, [ [\na_\mu , \qp] \, ,u_\nu ] 
 \na^\mu \hc~,\quad
 \, \langle   Q_+  \rangle  \, [ [v \cdot \na , \qp] \, , S \cdot u  ]~, \no  \\
&82\qquad& \gamma^\mu  \gamma_5  \, \langle [ \na_\mu ,\qp] \, [\qp , u_\nu  ] \rangle 
\na^\nu \hc~,  \quad 
 \, \langle [S \cdot \na ,\qp] \, [\qp , v \cdot u  ] \rangle~, \no \\
&83\qquad&  \gamma^\nu  \gamma_5  \, \langle [ \na_\mu ,\qp] \, [\qp , u_\nu  ] \rangle 
\na^\mu  \hc~, \quad 
\, \langle [v \cdot \na ,\qp] \, [\qp , S \cdot u  ] \rangle~, \no \\
&84\qquad& \gamma^\mu \gamma_5  \, \langle [\na_\mu ,\qm] \, [\qm , u_\nu  ] \rangle 
\na^\nu  \hc~, \quad
 \, \langle [ S\cdot \na ,\qm] \, [\qm , v \cdot u  ] \rangle~, \no\\
&85\qquad&  \gamma^\nu \gamma_5  \, \langle [\na_\mu ,\qm] \, [\qm , u_\nu  ] \rangle 
\na^\mu  \hc~, \quad
\, \langle [ v\cdot \na ,\qm] \, [\qm , S \cdot u  ] \rangle~, \no\\
&86\qquad& \gamma^\nu \gamma_5  \, \langle [\na_\mu , \qp] \, [\na_\nu , \qm ] \rangle 
i\na^\mu \hc~,\quad 
 \, \langle [i v \cdot \na , \qp] \, [ S \cdot \na , \qm ] \rangle~, \no \\
&87\qquad& \gamma^\mu \gamma_5  \, \langle [\na_\mu , \qp] \, [\na_\nu , \qm ] \rangle 
i\na^\nu  \hc~, \quad 
\, \langle [i S \cdot \na , \qp] \, [ v \cdot \na , \qm ] \rangle~, \no \\
&88\qquad& 
i \sig \gamma_5  \, \langle [\na_\mu , [\na_\nu , \qp]] \,  \qm \rangle~,\quad 
 \SIG  \, \langle [\na_\mu , [\na_\nu , \qp]] \,  \qm \rangle~, \no \\
&89\qquad&
i\sig \gamma_5  \, \langle [\na_\mu , [\na_\nu , \qm]] \,  \qp \rangle~, \quad 
 \SIG  \, \langle [\na_\mu , [\na_\nu , \qm]] \,  \qp \rangle~, \no \\
&90\qquad& 
i\sig \gamma_5  \, \langle Q_+  \rangle \,   [ \na_\mu , [ \na_\nu , \qm ]]~, 
\quad  \SIG  \, \langle Q_+  \rangle \,   [ \na_\mu , [ \na_\nu , \qm ]]~.
\eeqa


\pagebreak

\newpage

\noindent {\Large \bf Figures}

\vspace{+1.5cm}

\end{document}